\documentclass{aa} 
\usepackage[varg]{txfonts}
\usepackage{graphicx}
\usepackage{caption}
\usepackage{booktabs}

\usepackage{natbib}

\usepackage{hyperref}
\hypersetup{
  colorlinks = true, 
  urlcolor  = blue, 
  linkcolor = red,  
  citecolor = blue  
}
\bibpunct{(}{)}{;}{a}{}{,}

\def\simlt{\mathrel{\rlap{\lower 3pt\hbox{$\sim$}}\raise 2.0pt\hbox{$<$}}}
\def\simgt{\mathrel{\rlap{\lower 3pt\hbox{$\sim$}} \raise 2.0pt\hbox{$>$}}}

\usepackage{amstext}

\begin{document}

\title{Infrared-radio relation in the local Universe}
\titlerunning{Infrared-radio relation in the local universe}
\author{
     K. Tisani\'c  \inst{1,7}  \thanks{\emph{ktisanic@irb.hr}}
  \and G. De Zotti  \inst{2}  
  \and A. Amiri  \inst{3,4,5,6}  
  \and A. Khoram    \inst{5,6,8}
  \and S. Tavasoli   \inst{6}
  \and Z. Vidovi\'c-Tisani\'c \inst{9}
}

\institute{
    Department of Physics, 
    Faculty of Science, 
    University of Zagreb, 
    Bijeni\v{c}ka cesta 32, 
    10000 Zagreb, 
    Croatia
    \and INAF - Osservatorio Astronomico di Padova, Vicolo dell'Osservatorio 5, I-35122 Padova, Italy
    \and Dipartimento di Fisica e Astronomia, Universit\`a di Firenze, Via G. Sansone 1, 50019, Sesto Fiorentino (Firenze), Italy 
    \and INAF - Osservatorio Astrofisico di Arcetri, Largo E. Fermi 5, 50127, Firenze, Italy
    \and School of Astronomy, Institute for Research in Fundamental Sciences (IPM), PO Box 19395-5746 Tehran, Iran 
    \and Physics Department, Kharazmi University, Tehran, Iran 
    \and Ru{\dj}er Bo\v{s}kovi\'{c} Institute, Bijeni\v{c}ka cesta 54, 10000 Zagreb, Croatia
    \and Dipartimento di Fisica e Astronomia G. Galilei, Universit\`a degli Studi di Padova, I-35131 Padova, Italy
    \and University of Zagreb, Faculty of Mining, Geology and Petroleum Engineering, Zagreb, Croatia
}

\abstract {The Square Kilometer Array (SKA) is expected to detect high-redshift galaxies with
star formation rates (SFRs) up to two orders of magnitude lower than
\textit{Herschel} surveys and will thus boost the ability of radio astronomy to study extragalactic sources. The tight infrared-radio correlation offers the possibility of using radio emission as a dust-unobscured star formation diagnostic. However, the physics governing the link between radio emission and star formation is poorly understood, and recent studies have pointed to differences in the exact calibration required when radio is to be used as a star formation tracer. } {We improve the calibration of the relation of the local radio luminosity--{SFR} and to test whether there are nonlinearities in it.  } { We used a sample of Herschel Astrophysical Terahertz Large Area Survey (H-ATLAS) sources and investigated their radio luminosity, which was derived using {the NRAO VLA Sky Survey (NVSS)  and Faint Images of the Radio Sky at Twenty-cm (FIRST)} maps. We stacked the bins of infrared luminosity and SFR and accounted for bins with no detections in the stacked images using survival analysis fitting. This approach was tested using Monte Carlo simulations.} {After removing sources from the sample that have excess radio emission, which is indicative of nuclear radio activity, we found no deviations from linearity of the mean relations between radio luminosity and either SFR or infrared luminosity. 
} {We analyzed the link between radio emission and {SFR} or {infrared} luminosity using a local sample of star-forming galaxies without evidence of nuclear radio activity and found no deviations from linearity{, although our data are also consistent with the small nonlinearity reported by some recent analyses}. The normalizations of these relations are intermediate between those reported by earlier works.}
\maketitle

\section{Introduction}\label{sect:introduction}

\begin{figure*}
    \centering
    \includegraphics[width=\textwidth]{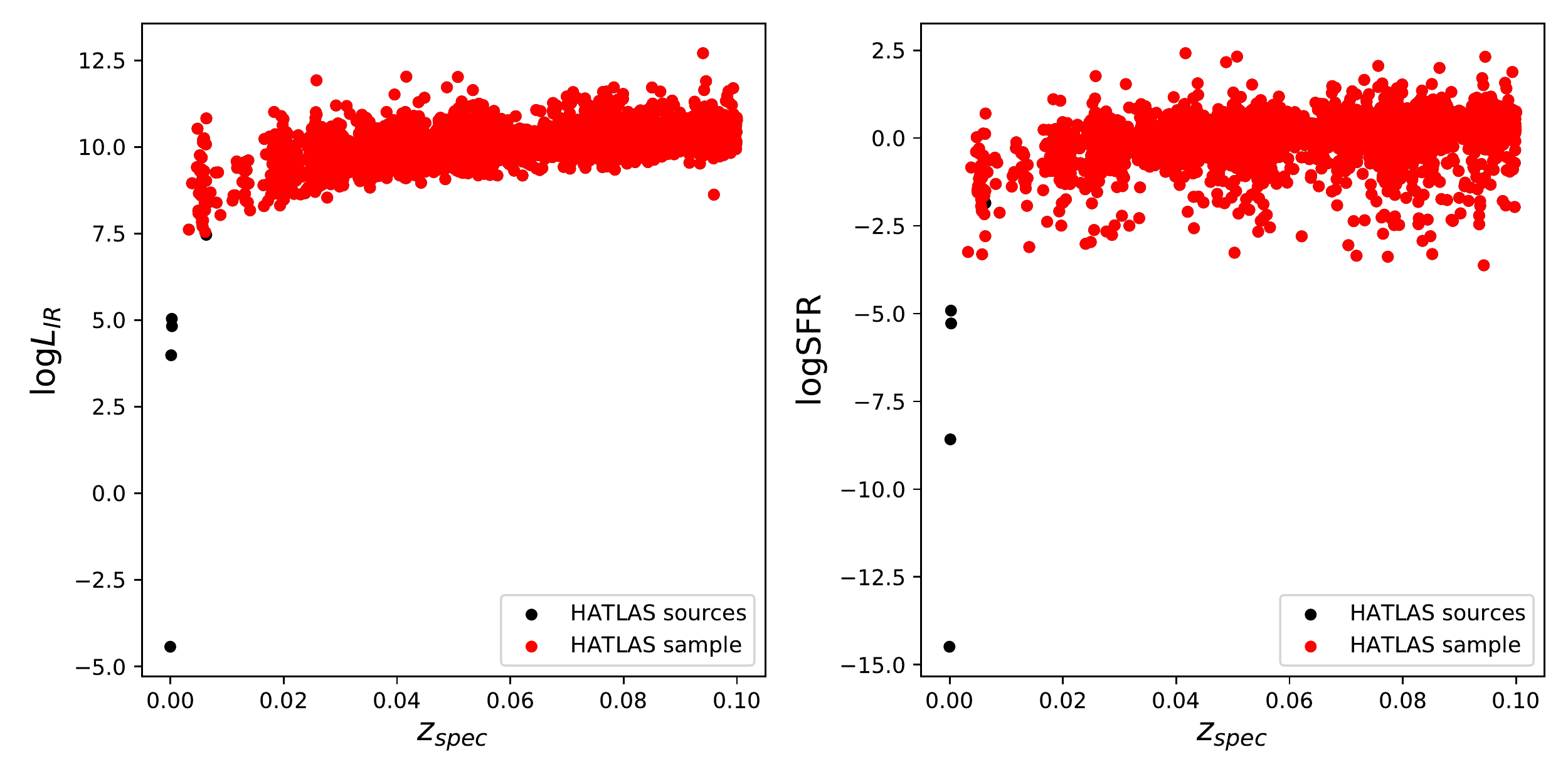}
    \caption{Derived infrared luminosity and star formation rate for all H-ATLAS sources and for the selected H-ATLAS sample. {The H-ATLAS sample (red points) is relatively complete in infrared luminosity and SFR and was used as a basis for the SFG subsample. }}
    \label{fig:Sample}
\end{figure*}

The discovery of a tight correlation between radio and far-infrared (FIR)
emission of star-forming galaxies {(SFGs)} \citep{deJong85, Helou85, Gavazzi1986,
Yun2001} has offered the possibility of using radio continuum emission as a
star formation tracer. Radio emission has two important advantages over
other star formation diagnostics \citep[for a review,
see][]{KennicuttEvans2012}: it is unaffected by dust extinction and measures
both the dust-obscured and the unobscured star formation. Unobscured star formation is
missed by FIR emission, and dust-obscured star formation is missed by {ultraviolet}, optical, and {near-infrared (NIR)} tracers.

The use of radio emission as a star formation diagnostic is
particularly important because the Square Kilometre Array (SKA) will
reach flux limits that are orders of magnitude fainter than is currently possible, and
it will achieve this over large areas. It will thus boost the potential of radio astronomical
observations for extragalactic studies, including for star formation history
\citep{Prandoni2015RevealingSurveys}.

The planned (SKA1-Mid) ultradeep survey at 1.4\,GHz, with a $5\,\sigma$
detection limit of $0.25\,\mu$Jy, is expected to detect high-$z$ galaxies with
star formation rates (SFRs) up to two orders of magnitude lower than
\textit{Herschel} surveys \citep{Mancuso2015PREDICTIONSGALAXIES}. This means
that SFRs well below those of typical {SFGs} up to redshifts
of 3--4 will be reached. At $z\simeq 1,$ this survey will detect galaxies down to SFRs of a few
$M_\odot\,\hbox{yr}^{-1}$ and will reach down to a $\hbox{few
hundred}\,M_\odot\,\hbox{yr}^{-1}$ at $z\simeq 10$.

An accurate determination of the relation between radio luminosity
and SFR or {infrared (IR)} luminosity provides one of the main criteria for separating radio
{active galactic nuclei (AGNs)} from SFGs in deep radio surveys: radio emission in
excess of the emission expected from SFR betrays the presence of radio activity of
nuclear origin.
However, the link between synchrotron emission, which dominates at low radio
frequencies, and star formation depends on complex and poorly understood
physics. As a consequence, the calibration, shape, and redshift
dependence of the radio luminosity-SFR relation are still
debated.

A linear relation between the 1.4\,GHz luminosity, $L_{1.4}$, and the SFR was
reported by several studies, based either on direct estimates of the SFR,
generally via the {Multi-wavelength Analysis of
Galaxy Physical Properties (MAGPHYS)} package \citep{daCunha2008}, or on the assumption
that the IR luminosity\footnote{Here by $L_{\rm IR}$ we mean the
luminosity integrated over the 8--$1000\,\mu$m range.} is a reliable SFR
measure \citep{Yun01, Murphy2011CalibratingNGC6946, Murphy2012TheRegions,
CalistroRivera2017The0z2.5, Delhaize17, Wang2019ATracer}.

There are significant differences among the calibrations reported by
different groups for the same initial mass function (IMF), however. For a
\citet{Chabrier03} IMF, the values of $\log(L_{1.4}/\hbox{W}\,\hbox{Hz}^{-1}) -
\log(\hbox{SFR}/M_\odot\,\hbox{yr}^{-1})$ at $z=0$ range from $20.96\pm 0.03$
\citep{Delhaize17} to $21.39\pm 0.04$ \citep{CalistroRivera2017The0z2.5}. This is a difference of a
factor of $\simeq 2.7$  in the $L_{1.4}$/SFR ratio\footnote{Whenever
the results are presented in terms of the best-fit value of $q_{\rm
IR}=\log(L_{\rm IR})-\log(L_{1.4})-\log(3.75\times 10^{12}),$ we have computed
the $L_{1.4}$/SFR ratio using the $L_{\rm IR}$/SFR calibration by
\citet{KennicuttEvans2012}, yielding the relation
$\log(L_{1.4}/\hbox{W}\,\hbox{Hz}^{-1}) -
\log(\hbox{SFR}/M_\odot\,\hbox{yr}^{-1})=23.836-q_{\rm IR}$.}.  Intermediate
values were found by \citet[][2.13, for
$T_e=10^4\,\hbox{K}$]{Murphy2011CalibratingNGC6946},
\citet[][$21.215^{+0.016}_{-0.013}$]{Sargent10}, \citet[][$21.21\pm
0.08$]{Magnelli15} , and \citet[][21.23]{Wang2019ATracer}.

A linear relation is consistent with the prediction of the calorimetric model
\citep{Voelk89}, according to which relativistic electrons produced by supernova
explosions lose all their energy inside the galaxy. On the other hand,
\citet{Bell2003EstimatingCorrelation} argued that the $L_{1.4}$/SFR ratio of
dwarf galaxies must be lower than that of the bright galaxies. A suppression of
synchrotron emission of small galaxies was indeed predicted
\citep{ChiWolfendale1990} because they are unable to confine relativistic
electrons, which can then escape into the intergalactic medium before releasing
their energy. Early observational evidence of a nonlinear radio-IR relation,
$L_{\rm radio}\propto L_{\rm IR}^\delta$ with $\delta > 1$, were presented by
\citet{Klein1984}, \citet{DevereuxEales1989}, and \citet{PriceDuric1992},
although this view was controversial \citep{Condon92}. In contrast, \citet{Gurkan2018LOFAR/H-ATLAS:Relation} found evidence of a substantial flattening of the $L_{\rm radio}$--SFR relation below $\hbox{SFR}\simeq 1\,M_\odot\,\hbox{yr}^{-1}$, with the slope decreasing from $\simeq 1$ to $\simeq 0.5${, while \citet{Matthews2021} estimated $\delta$ to be $\sim0.85$.} The substantial excess radio emission of galaxies with low SFR with respect to extrapolations from galaxies with high SFRs was interpreted as an indication of an additional mechanism that was thought to operate in low-SFR galaxies to generate radio-emitting relativistic electrons.

Tests of the local relations between $L_{1.4}$ and the SFR can be made
by comparing the observationally determined local SFR function
\citep{Mancuso2015PREDICTIONSGALAXIES, Aversa2015} with the local radio
luminosity function of SFGs \citep{MauchSadler2007}. For consistency, it is
necessary to assume a decline in $L_{\rm sync}$/SFR ratio with decreasing SFR  \citep{Massardi2010, Bonato2017}.

On the other hand, for $L_{\rm IR}\simlt 10^{10}\,L_\odot$, the infrared
luminosity is no longer a reliable proxy of the SFR \citep{Clemens2010}. The
standard calibration can either lead to an overestimate of the SFR if $L_{\rm
IR}$ is dominated by dust that is heated by old stellar populations, or to an
underestimate because it may miss the unabsorbed emission by young stars.

This synthetic review shows that several important questions are still
open. These include the questions to which extent the $L_{\rm IR}$ is a reliable SFR tracer,
what shape and normalization hold for the $L_{\rm IR}$-SFR relation, whether
the $L_{1.4}$--SFR relation is linear, and  which is the best value of the
coefficient of this relation.

\begin{figure}
    \includegraphics[width=\columnwidth]{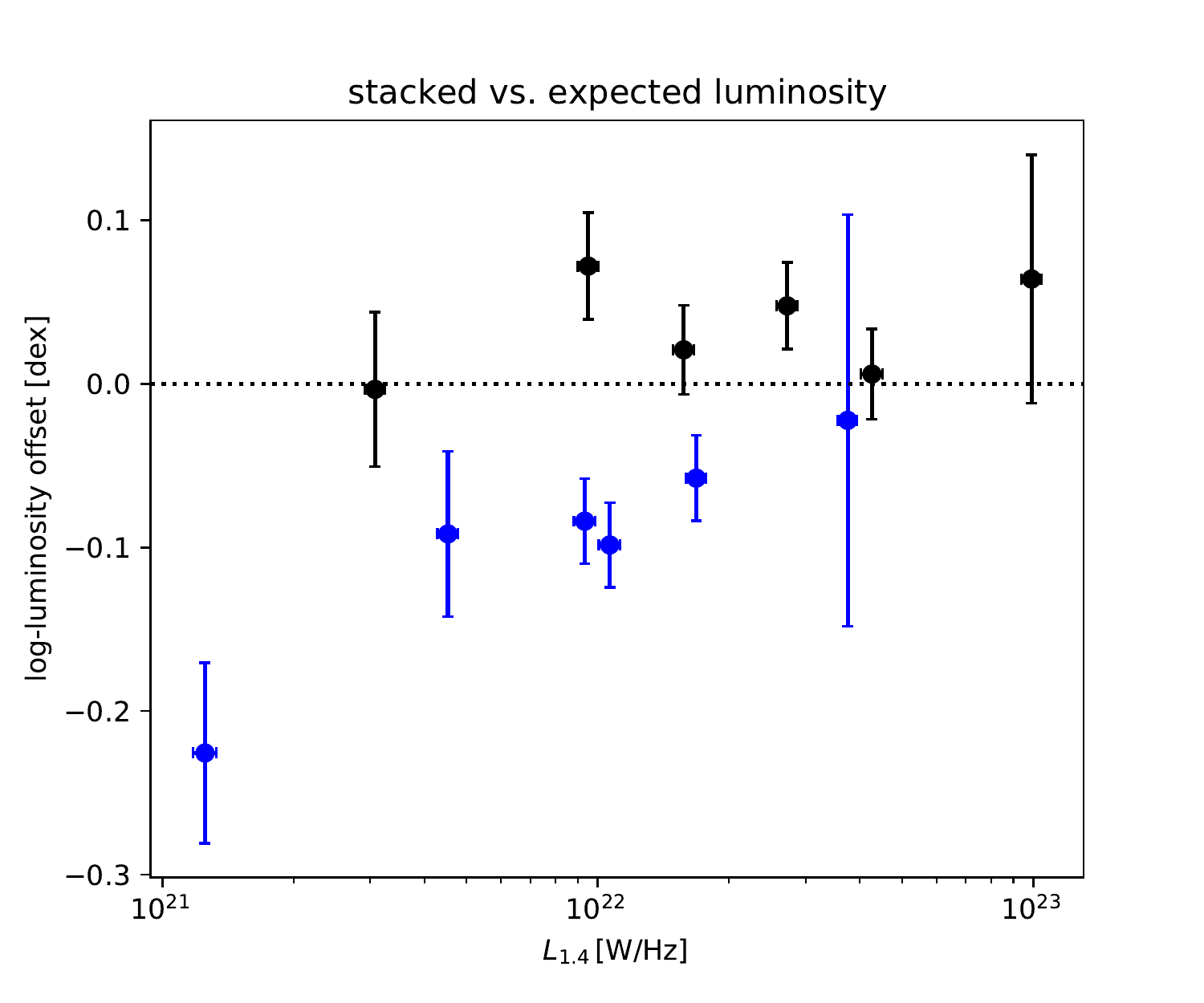}
    \caption{Logarithmic offsets between $1.4\,\mathrm{GHz}$ radio luminosities derived from stacked cutouts using BLOBCAT compared to the values derived from catalog flux densities. The black (blue) points refer to the subset of the H-ATLAS sample detected by the NVSS (FIRST) surveys.}
    \label{fig:Correction}
\end{figure}

\begin{figure*}
    \centering
     \includegraphics[width=.85\textwidth]{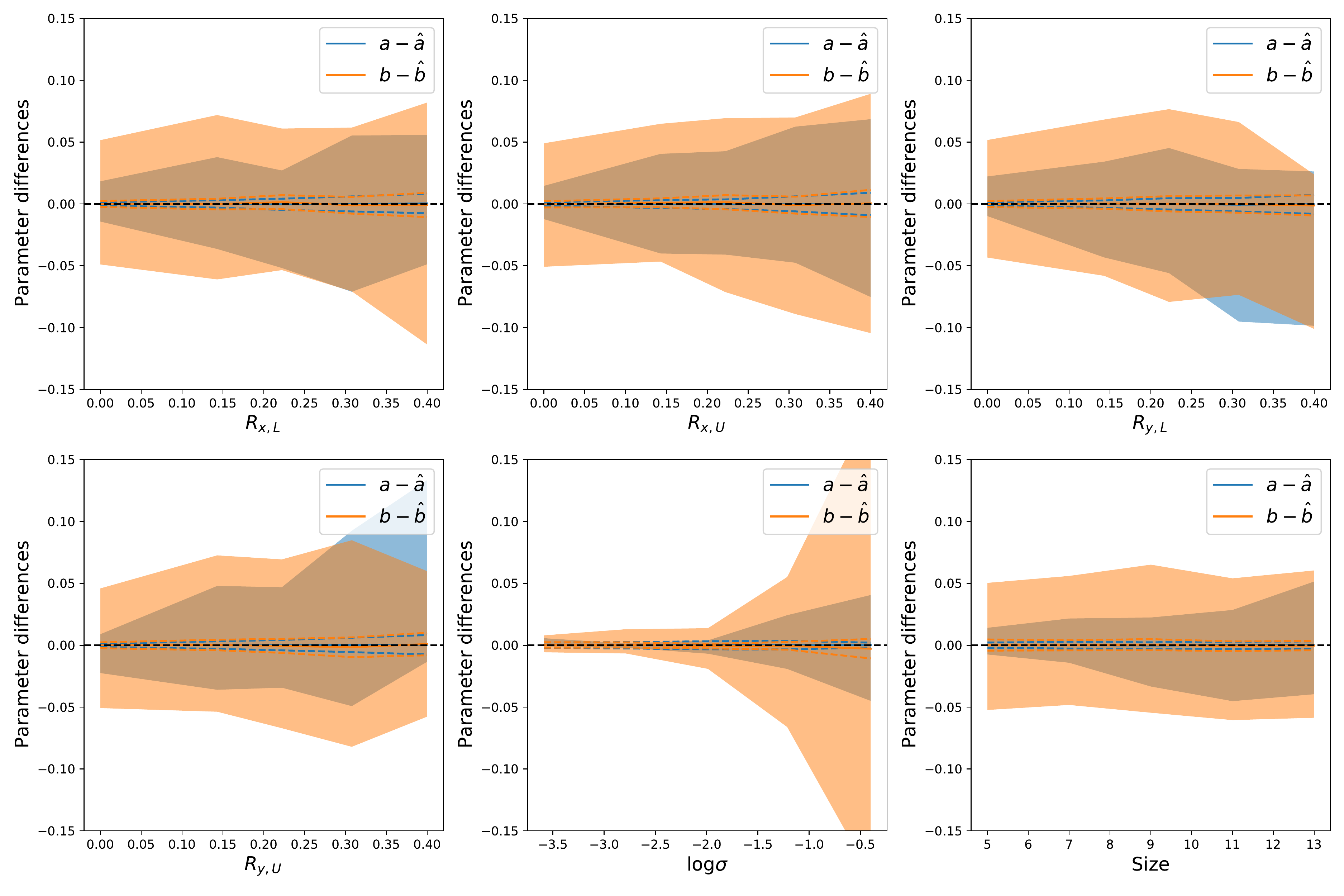}
    \caption{Differences between the Monte Carlo simulated linear model parameters and the corresponding simulation inputs. The results are shown for bins in the ratio of {the number of} upper and lower limits on the $x$ - and $y$ -axes ($R_{x,L}$, $R_{x,U}$, $R_{y,L}$ ,and $R_{y,U}$) as well as for bins of the dispersion $\sigma$ and of the size of the sample (see Sect. \ref{sect:survival} for details). The blue lines show the median differences between the simulated, $a$, and input slopes, $\hat a$, and the orange lines show the same difference for the simulated intercept, $b,$ and its input in the simulation, $\hat b$. The shaded intervals are the intervals between the $16\text{th}^{\mathrm{}}$ and $84\text{th}^{\mathrm{}}$ percentile in each bin. }\label{fig:FigureSimulationDiff}
    \end{figure*}
    \begin{figure*}
    \centering
    \includegraphics[width=.85\textwidth]{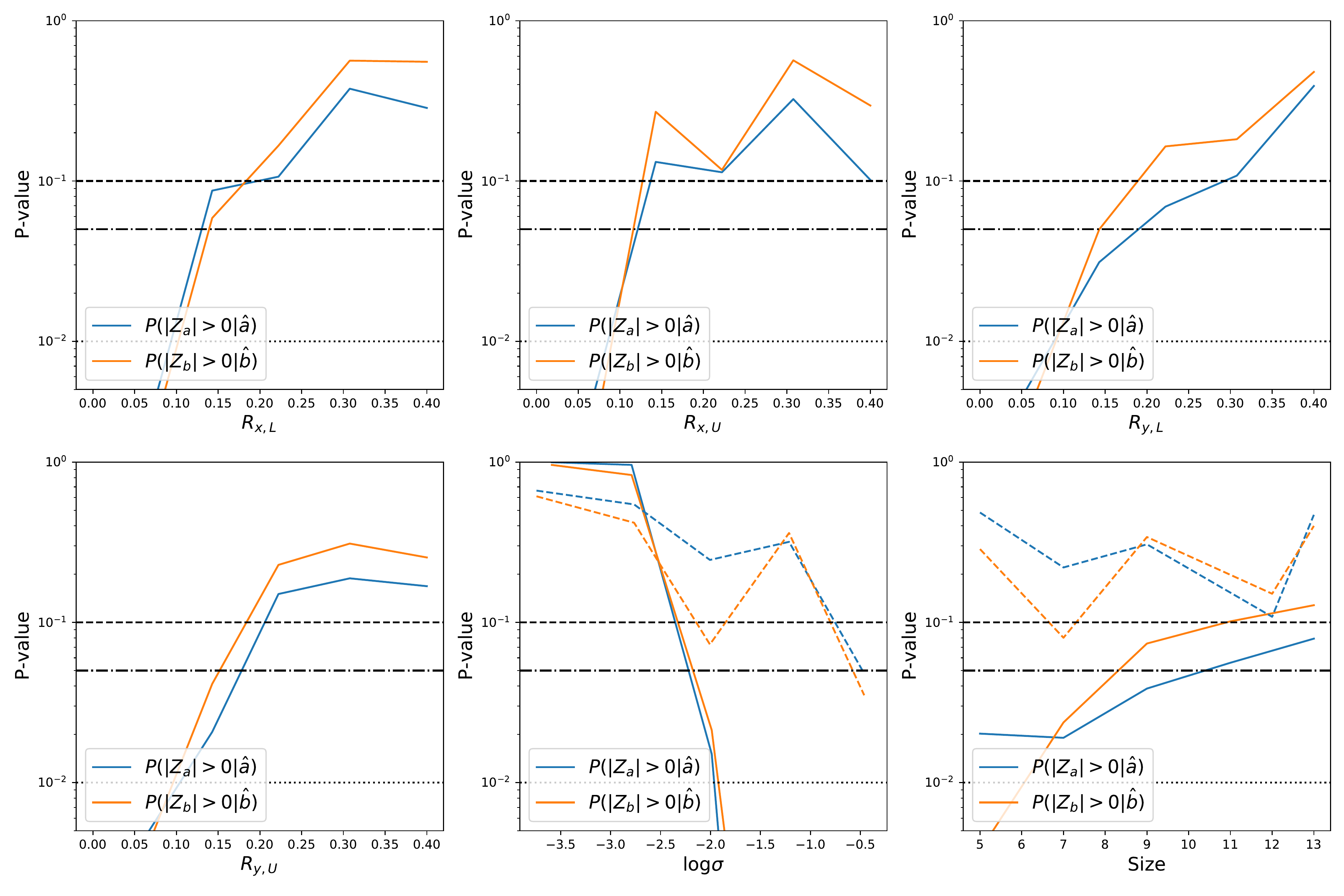}
    \caption{P-values for the differences between the Monte Carlo simulated linear model parameters and their corresponding simulation inputs. The blue lines show the P-values computed using the Z-scores of the differences between the simulated, $a$, and input slopes, $\hat a$, and the orange lines show the P-values computed for the simulated intercept. The bins are the same as in Fig. \ref{fig:FigureSimulationDiff} (see Sect. \ref{sect:survival} for details). The dashed, dot-dashed, and dotted black lines show the $10\%$, $5\%$, and $1\%$ confidence limits, respectively, as reference. }\label{fig:FigureSimulationPvalues}
    \end{figure*}
    
To answer these questions, we need solid determinations of $L_{\rm IR}$ and of
the SFR as well as 1.4\,GHz measurements. In this paper, we focus on nearby
galaxies and take advantage of the wealth of multifrequency data for the three
equatorial fields of the \textit{Herschel} Astrophysical Terahertz Large Area Survey
\citep[H-ATLAS;][]{Eales2010}. This is the largest extragalactic survey and was carried out
with the \textit{Herschel} Space Observatory \citep{Pilbratt2010}. Each of the equatorial fields covers an area of about $54\,\hbox{deg}^2$, centered at approximately 9, 12, and 15\,h in right ascension \citep{Valiante2016}.

The sample is described in Sect.~\ref{sect:sample} and the method in
Sect.~\ref{sect:method}. The results are presented in Sect.~\ref{sect:results} {and are discussed in Sect.~\ref{sect:discussion}. In} Sect.~\ref{sect:conclusions} we summarize our main conclusions. We
adopt a flat {$\Lambda$ cold dark matter ($\Lambda$CDM)} cosmology with
$H_0=67.37\,\hbox{km}\,\hbox{s}^{-1}\,\hbox{Mpc}^{-1}$ and $\Omega_m = 0.315$
\citep{Planck2018parameters}.

\section{Data and sample}\label{sect:sample}

\begin{table*}\scriptsize
\setlength{\tabcolsep}{3pt}
    \centering
    \caption{Number of sources in the chosen sample within  $\log\mathrm{L_{\rm IR}}$ or $\log\mathrm{SFR}$ bins. }
    \label{tab:Number_counts_table}
    \begin{tabular}{c c c c c c c c c c c c c c c c c c c c c c c c}
\toprule
\toprule
Bin & \multicolumn{7}{c}{H-ATLAS sample}  &  \multicolumn{4}{c}{SFG $5\sigma$ subsample} &  \multicolumn{4}{c}{SFG $3\sigma$ subsample} \\
\cmidrule{2-8}\cmidrule{13-16}
$\log\mathrm{L_{\rm IR}}$ & Total  & \multicolumn{2}{c}{Detections} & \multicolumn{2}{c}{In $5\sigma$ stack} & \multicolumn{2}{c}{In $3\sigma$ stack} &  \multicolumn{2}{c}{FIRST subsample}   & \multicolumn{2}{c}{NVSS subsample}    
& \multicolumn{2}{c}{FIRST subsample}   & \multicolumn{2}{c}{NVSS subsample}     \\ 
  & &    FIRST& NVSS & FIRST& NVSS & FIRST& NVSS & Total (detections)  & In stack   & Total (detections)  & In  stack  
& Total (detections)  & In stack   & Total (detections)  & In stack  \\ 
\midrule
$\langle 7.5, 8.0]$           &   6 & 0&0      & N& N & N& N     & 4 (0)      & N & 0 (0)      & N   & 4 (0)      & N & 0 (0)      & N  \\
$\langle 8.0, 8.5]$           &  16 & 0&0      & N& N & N& Y     & 8 (0)      & N & 2 (0)      & N   & 8 (0)      & N & 2 (0)      & N  \\
$\langle 8.5, 8.8]$           &  27 & 0&1      & N& N & N& N     & 17 (0)     & N & 8 (1)      & N   & 17 (0)     & Y & 8 (1)      & N  \\
$\langle 8.8, 9.0]$           &  37 & 1&1      & N& N & N& N     & 24 (1)     & Y & 5 (1)      & N   & 24 (1)     & Y & 5 (1)      & N  \\
$\langle 9.0, 9.2]$           &  54 & 1&0      & N& N & N& Y     & 33 (0)     & N & 5 (0)      & N   & 34 (0)     & N & 5 (0)      & N  \\
$\langle 9.2, 9.4]$           & 111 & 0&0      & N& Y & Y& Y     & 69 (0)     & Y & 13 (0)     & Y   & 69 (0)     & Y & 13 (0)     & Y  \\
$\langle 9.4, 9.6]$           & 189 & 1&4      & N& Y & Y& Y     & 101 (1)    & N & 24 (2)     & N   & 101 (1)    & Y & 24 (2)     & Y  \\
$\langle 9.6, 9.8]$           & 280 & 1&2      & N& N & Y& Y     & 155 (1)    & Y & 28 (1)     & N   & 155 (1)    & Y & 28 (1)     & Y  \\
$\langle 9.8, 9.9]$           & 219 & 0&3      & N& N & Y& N     & 103 (0)    & Y & 12 (1)     & N   & 103 (0)    & Y & 12 (1)     & Y  \\
$\langle 9.9, 10.0]$          & 267 & 0&1      & N& Y & Y& Y     & 150 (0)    & Y & 25 (1)     & N   & 150 (0)    & Y & 26 (1)     & Y  \\
$\langle 10.0, 10.1]$         & 291 & 3&4      & Y& Y & Y& Y     & 166 (2)    & Y & 18 (1)     & Y   & 167 (2)    & Y & 21 (2)     & Y  \\
$\langle 10.1, 10.2]$         & 295 & 3&9      & Y& Y & Y& Y     & 193 (2)    & Y & 30 (4)     & Y   & 194 (2)    & Y & 35 (6)     & Y  \\
$\langle 10.2, 10.3]$         & 317 & 5&7      & Y& Y & Y& Y     & 238 (5)    & Y & 26 (6)     & Y   & 238 (5)    & Y & 28 (6)     & Y  \\
$\langle 10.3, 10.4]$         & 317 & 9&12     & Y& Y & Y& Y     & 264 (9)    & Y & 44 (11)    & Y   & 264 (9)    & Y & 45 (11)    & Y  \\
$\langle 10.4, 10.5]$         & 256 & 5&6      & Y& Y & Y& Y     & 226 (5)    & Y & 40 (1)     & Y   & 226 (5)    & Y & 42 (1)     & Y  \\
$\langle 10.5, 11.5]$         & 628 & 53&79    & Y& Y & Y& Y     & 599 (53)   & Y & 257 (65)   & Y   & 595 (51)   & Y & 262 (64)   & Y  \\
$\langle 11.5, 12.0]$         &  15 & 2&9      & Y& Y & Y& Y     & 15 (2)     & Y & 13 (9)     & Y   & 15 (2)     & Y & 14 (9)     & Y  \\
$\langle 12.0, 13.0]$         &   3 & 0&0      & N& N & N& N     & 3 (0)      & N & 3 (0)      & N   & 3 (0)      & N & 3 (0)      & N  \\
Total                         &3328 & 84& 138  & 7 &   10 & 12 &      13 & 2368 ( 81) &  12 & 553 (104)  &   8  & 2367 ( 79) &  14 & 573 (106)  & 12  \\
\midrule
$\log\mathrm{SFR}$ \\
\midrule
$\langle -3.7, -2.5]$         &  22 & 1&0     &  N& N & N& Y     & 0 (0)      & N & 0 (0)      & N   & 0 (0)      & N & 0 (0)      & N \\
$\langle -2.5, -2.0]$         &  33 & 1&1     &  N& N & N& N     & 2 (0)      & N & 0 (0)      & N   & 2 (0)      & N & 0 (0)      & N \\
$\langle -2.0, -1.5]$         &  45 & 0&1     &  N& Y & N& Y     & 4 (0)      & N & 0 (0)      & N   & 5 (0)      & N & 0 (0)      & N \\
$\langle -1.5, -1.0]$         & 108 & 0&3     &  Y& N & Y& N     & 22 (0)     & N & 8 (0)      & N   & 23 (0)     & Y & 10 (1)     & N \\
$\langle -1.0, -0.6]$         & 191 & 1&2     &  Y& Y & Y& Y     & 70 (1)     & Y & 11 (2)     & Y   & 71 (1)     & Y & 11 (2)     & Y \\
$\langle -0.6, -0.2]$         & 478 & 5&7     &  Y& Y & Y& Y     & 224 (5)    & Y & 38 (3)     & Y   & 225 (5)    & Y & 42 (4)     & Y \\
$\langle -0.2, -0.1]$         & 252 & 2&4     &  Y& Y & Y& Y     & 149 (2)    & Y & 24 (2)     & Y   & 149 (2)    & Y & 26 (2)     & Y \\
$\langle -0.1, 0.1]$          & 524 & 0&6     &  Y& Y & Y& Y     & 303 (0)    & Y & 42 (3)     & Y   & 303 (0)    & Y & 44 (4)     & Y \\
$\langle 0.1, 0.2]$           & 303 & 5&6     &  Y& Y & Y& Y     & 243 (4)    & Y & 35 (5)     & Y   & 242 (4)    & Y & 35 (5)     & Y \\
$\langle 0.2, 0.3]$           & 353 & 9&9     &  Y& Y & Y& Y     & 339 (9)    & Y & 40 (6)     & Y   & 339 (9)    & Y & 42 (6)     & Y \\
$\langle 0.3, 0.4]$           & 245 & 6&9     &  Y& Y & Y& Y     & 242 (6)    & Y & 40 (6)     & Y   & 240 (5)    & Y & 44 (6)     & Y \\
$\langle 0.4, 0.6]$           & 406 & 18&24   &  Y& Y & Y& Y     & 403 (18)   & Y & 81 (16)    & Y   & 402 (17)   & Y & 83 (16)    & Y \\
$\langle 0.6, 2.5]$           & 368 & 36&66   &  Y& Y & Y& Y     & 367 (36)   & Y & 234 (61)   & Y   & 366 (36)   & Y & 236 (60)   & Y \\
Total                         &3328 & 84& 138 &   10 & 10  & 10 &    11   & 2368 ( 81) &  9 & 553 (104)  &   9  & 2367 ( 79) & 10  & 573 (106)  &  9 \\
\bottomrule
    \end{tabular}
    \tablefoot{The numbers in parentheses indicate {sources individually detected}  by the FIRST and NVSS surveys. {The SFG subsamples contain sources satisfying the \citet{Smolcic:17b} criterion applied using FIRST and NVSS derived radio luminosities or their $5\sigma$ or $3\sigma$ upper limits. The `Total' columns show the total number of sources in each (sub)sample, the `in stack' columns show whether a particular bin was processed as a detection (Y) or an upper limit (N) in a particular radio map, out to the detection limit of either 5 or 3 $\sigma$. In the `Total' rows, the number of Ys is shown for the appropriate columns.}}
\end{table*}

In addition to the \textit{Herschel} photometry in five bands ($100$, $160$,
$250$, $350$, and $500\,\mathrm{\mu m}$), the H-ATLAS equatorial fields have
been imaged in the {NIR and mid-infrared (MIR)} with the Wide-field Infrared Survey Explorer
\citep[WISE;][]{Wright2010}, in {the NIR and optical} with the UK Infrared Deep
Sky Survey Large Area Survey \citep[UKIDSS-LAS;][]{Lawrence2007}, with the
VISTA Kilo-Degree Infrared Galaxy Survey \citep[VIKING;][]{Edge2013} and with
the VLT Survey Telescope (VST) Kilo-Degree Survey \citep[KIDS;][]{deJong2013}, and
in the {ultraviolet} with the Galaxy Evolution Explorer \citep[GALEX;][]{Martin2005}.
Redshifts were provided by the Sloan Digital Sky Survey
\citep[SDSS;][]{Abazajian2009}, the Galaxy and Mass Assembly survey
\citep[GAMA;][]{Driver2009, Liske2015}, and the 2-Degree-Field Galaxy Redshift
Survey \citep[2dF;][]{Colless2001}.

The three fields called GAMA9, GAMA12, and GAMA15, whose numbers indicate
the approximate right ascension of their centers, have areas of
$53.46\,\mathrm{deg^2}$, $53.56\,\mathrm{deg^2}$ , and $54.56\,\mathrm{deg^2}$,
respectively. The $4\,\sigma$ detection limit at the selection wavelength
($250\,\mu$m) is approximately 29.4\,mJy, and the catalog is complete to $\simeq 90\%$
 \citep{Valiante2016}. Corrections for flux density biases, obtained
by injecting artificial sources into the images, and aperture corrections were applied.

Reliable optical counterparts have been determined by \citet{Bourne2016}. We selected galaxies with spectroscopic redshift $z\le 0.1$ and
$\hbox{F250\_best} \geq 30\,$mJy, further requiring $\hbox{z\_quality} \geq 3$
(reliable redshifts) and $\hbox{GSQ-flag}=0$ (to skip stars and QSOs).
According to \citet{Bourne2016}, the completeness of the $z\le 0.1$ sample is
$91.3\pm 2.2$ \%.

This selection yielded 3,333 galaxies. For each of them, we evaluated the
8--$1000\,\mu$m luminosity, $L_{\rm IR}$, and the SFR. We used all the
photometric data. They were processed with the MAGPHYS package \citep{daCunha08} as updated by
\citet{Berta13}. MAGPHYS ensures consistent modeling of the observed spectral energy
distribution from {ultraviolet} to millimeter wavelengths. The data for our objects allow a good sampling of the full spectral energy distribution, and hence a reliable determination of both $L_{\rm IR}$ and of the SFR. 
In Fig.~\ref{fig:Sample} we show the derived {IR}
luminosities and SFRs as a function of redshift.

Radio data are provided by the NRAO VLA Sky Survey
\citep[NVSS;][]{Condon1998NVSS} and by the {Faint Images of the Radio Sky at Twenty-cm} \citep[FIRST;][]{Becker1994}, both at 1.4\,GHz. The NVSS
covered $82\%$ of the celestial sphere with a resolution of $45''$ and a
typical {root mean square (RMS)} noise of $0.45\,\mathrm{mJy/beam}$. The FIRST survey covered over
$10,000\,\hbox{deg}^2$ {($24\%$ of the celestial sphere)} at higher resolution \citep[$5.4''$;][]{Helfand2015} and
a typical {RMS} of 0.15 mJy/beam.

A fraction of the radio luminosity in our sample might be of nuclear origin. To select SFGs as opposed to radio-loud active galactic nuclei (RL AGNs), we applied the \citet{Smolcic:17b} $3\sigma$ radio excess criterion, according to which SFGs have $L_{\rm 1.4GHz}/\hbox{SFR}$ ratios below the redshift-dependent threshold,
\begin{equation}
\log\left(\frac{L_{1.4}}{\hbox{SFR}}\right)=21.984\times(1+z)^{0.013},    
\end{equation}
with $L_{1.4}$ in $\hbox{W}\,\hbox{Hz}^{-1}$ and SFR in $M_{\odot}\,\hbox{yr}^{-1}$. Sources below this threshold form the SFG subsample. 

If a source is not detected by the NVSS or the FIRST survey, we computed the $5\,\sigma$ local {RMS} noise of the cutouts converted into radio luminosity (see Sect. \ref{sect:method}) and applied the \citet{Smolcic:17b} criterion.  
{The SFG classification obtained in this way is conservative in the sense that it selects against AGN while possibly removing SFGs with a low SFR.}

{Our sample is not homogeneous in luminosity and redshift for the faintest sources, which, moreover, have unreliable distances. We therefore limited ourselves} to 
{$z_{\rm spec}>0.001$ and $\log(L_{\rm IR}/L_\odot)>7.5$. Hereafter we refer to this subset as the H-ATLAS sample. This sample (red points in Fig.\,\ref{fig:Sample}) contains 3,328
sources.} {The SFG subsamples depend on the \citet{Smolcic:17b} criterion and therefore depend on the radio map that was used to compute the radio luminosities. The SFG subsample based on FIRST radio luminosities contains 2368 sources, and the sample based on NVSS radio luminosities yields 553 sources.}

{To maximize the ratio of detections versus nondetections in stacks of radio maps, we defined extended subsamples for which we lowered the radio detection limit to $3\,\sigma$ in the stacking procedure described in Sect. \ref{sect:stacking}.  We label the results when the $3\sigma$ radio detection limit was used as the H-ATLAS $3\sigma$ sample and the corresponding SFG $3\sigma$ subsamples as the SFG $3\sigma$ samples. The SFG $3\sigma$ subsamples depend on the \citet{Smolcic:17b} criterion and therefore depend on the radio map that was used to compute the radio luminosities. The SFG $3\sigma$ subsample based on the FIRST radio luminosities contains 2367 sources, and the sample based on NVSS radio luminosities contains 573 sources.} The number of sources in different bins for each sample and subsample we defined above is given in Table\,\ref{tab:Number_counts_table}.

\section{Methods}\label{sect:method}

We used median radio luminosity stacking to account for nondetections in the sample. In Sect. \ref{sect:stacking} we describe the stacking method and test for possible biases when NVSS and FIRST maps are used, while in Sect. \ref{sect:survival} we describe the survival analysis method we used to estimate the impact of stacked images without detections.
\subsection{Stacking procedure}\label{sect:stacking}
Median stacking on the FIRST and NVSS maps was performed using cutouts centered on sources in each bin. 
These cutouts were then converted into $1.4\,\mathrm{GHz}$ radio luminosity density using the following equation:
\begin{equation}
L_{1.4} = \frac{4\pi D_L^2}{(1+z)^{1-\alpha}} S_o,
\end{equation}
where $D_L$ is the luminosity distance, $z$ is the spectroscopic redshift, $S_o$ is the observed flux density, and $\alpha$ is the assumed spectral index, $\alpha=0.7$ ($S_\nu\propto \nu^{-\alpha}$).

To ensure proper positioning of cutouts, each of them was centered using a cubic convolution interpolation to the source position in the H-ATLAS catalog. {After centering, stacked cutouts were produced by combining each pixel separately using median stacking.} { The $41\times 41\,\mathrm{px^2}$ stacked cutout and the $5\,\sigma$ noise estimate were used to infer the stacked radio luminosity densities using the source-extraction software BLOBCAT, which is designed to process radio-wavelength images \citep{Hales12}. The RMS noise estimate that we used as input for BLOBCAT was computed excluding the central circular region ($10\,\mathrm{px}$ radius) in the stacked cutout.}  The cutouts were limited in size to $41\,\mathrm{px}$ in each dimension in order to have a wide enough region in each cutout for RMS noise estimation without having to switch to the adjacent map for any source in the H-ATLAS sample.

The NVSS and FIRST maps have different resolutions. The $41\times41\,\mathrm{pix}^2$ area corresponds to  $630\arcsec\times630\arcsec$ and $75\arcsec\times75\arcsec$, respectively. 
{We considered an H-ATLAS source to have a valid radio counterpart if the position in the FIRST (NVSS) catalog was within a $10\arcsec$ radius from the H-ATLAS position. The number of detected sources is shown for each bin in Table\,\ref{tab:Number_counts_table}. } 

We tested whether radio luminosities derived using BLOBCAT may be affected by a systematic offset such as the snapshot bias \citep{White2007SignalsSurvey}. To this end, we compared them to luminosities derived from cataloged flux densities of detected sources.  {As shown in Fig. \ref{fig:Correction}, } we found that the offsets in both maps are within $\pm 0.1 \,\mathrm{dex}$, and only the FIRST stacks show a significant ($2\sigma$) negative trend with radio luminosity.  {This trend, however, is not sufficiently large to explain the differences between the quantities derived in Sect. \ref{sect:results} when the FIRST and NVSS maps are used.} {We verified whether the inclusion of corrections might change the results of Sect. \ref{sect:results}. Because no significant differences were found, no corrections were applied to the radio luminosity densities reported in this paper.}

\subsection{Survival analysis}\label{sect:survival}

Stacking yielded $5\,\sigma$ detections for most but not all bins of the H-ATLAS and the SFG subsamples, as summarized in Table~\ref{tab:Number_counts_table}. A further complication is that various quantities of interest might have upper or lower limits in the two regression axes. {An example of an upper limit is the $L_{1.4}$ nondetection, while an example of a lower limit is the $q_{IR}$ nondetection due to the $L_{1.4}$ upper limit.} For these cases, we extended the \citet{Sawicki2012SEDfit:Data} method to the error-in-variable case by deriving its total least-squares variant. {We estimated the flux upper limits as $n$ times the RMS noise of the stacked cutout, with $n=5$ for the $5\sigma$ BLOBCAT detection threshold and $n=3$ for the $3\sigma$ threshold.}

We modeled a dataset as a pair of values $(x_i,y_i)$ drawn from normal distributions with variances $\sigma_{x,i}^2$ and $\sigma_{y,i}^2$. For upper limits, we assumed $x\leq x_i^U$ or $y\leq y_i^U$ , and for lower limits, we assumed $x\geq x_i^L$ or $y\geq y_i^L$. For a general model $f(x,\theta),$ we adopted the likelihood function 
\begin{equation}
\begin{aligned}
    L(\theta)&=\prod\limits_{i\in D_x\cap D_y}\frac{1}{\sigma_{x,i}\sigma_{y,i}}\phi\left(\frac{X_i-x_i}{\sigma_{x,i}}\right)\phi\left(\frac{Y_i-f(\theta,X_i)}{\sigma_{y,i}}\right)\\
    &\times\prod\limits_{i\in D_x\cap U_y}\frac{1}{\sigma_{x,i}\sigma_{y,i}}\phi\left(\frac{X_i-x_i}{\sigma_{x,i}}\right)\Phi\left(\frac{y_i^U-f(\theta,X_i)}{\sigma_{y,i}}\right)\\
    &\times\prod\limits_{i\in D_x\cap L_y}\frac{1}{\sigma_{x,i}\sigma_{y,i}}\phi\left(\frac{X_i-x_i}{\sigma_{x,i}}\right)\left(1-\Phi\left(\frac{y_i^L-f(\theta,X_i)}{\sigma_{y,i}}\right)\right)\\
    &\times\prod\limits_{i\in U_x\cap D_y}\frac{1}{\sigma_{x,i}\sigma_{y,i}}\int\limits_{-\infty}^{x_i^U}dX_i\phi\left(\frac{X_i-x_i}{\sigma_{x,i}}\right)\phi\left(\frac{Y_i-f(\theta,X_i)}{\sigma_{y,i}}\right)\\
    &\times\prod\limits_{i\in L_x\cap D_y}\frac{1}{\sigma_{x,i}\sigma_{y,i}}\int\limits_{x_i^L}^{\infty}dX_i\phi\left(\frac{X_i-x_i}{\sigma_{x,i}}\right)\phi\left(\frac{Y_i-f(\theta,X_i)}{\sigma_{y,i}}\right)\\
    &\times\prod\limits_{i\in U_x\cap U_y}\frac{1}{\sigma_{x,i}\sigma_{y,i}}\int\limits_{-\infty}^{x_i^U}dX_i\int\limits_{-\infty}^{y_i^U}dY_i\phi\left(\frac{X_i-x_i}{\sigma_{x,i}}\right)\phi\left(\frac{Y_i-f(\theta,X_i)}{\sigma_{y,i}}\right)\\
    &\times\prod\limits_{i\in U_x\cap L_y}\frac{1}{\sigma_{x,i}\sigma_{y,i}}\int\limits_{-\infty}^{x_i^U}dX_i\int\limits_{y_i^L}^{\infty}dY_i\phi\left(\frac{X_i-x_i}{\sigma_{x,i}}\right)\phi\left(\frac{Y_i-f(\theta,X_i)}{\sigma_{y,i}}\right)\\
    &\times\prod\limits_{i\in L_x\cap U_y}\frac{1}{\sigma_{x,i}\sigma_{y,i}}\int\limits_{x_i^L}^{\infty}dX_i\int\limits_{-\infty}^{y_i^U}dY_i\phi\left(\frac{X_i-x_i}{\sigma_{x,i}}\right)\phi\left(\frac{Y_i-f(\theta,X_i)}{\sigma_{y,i}}\right)\\
    &\times\prod\limits_{i\in L_x\cap L_y}\frac{1}{\sigma_{x,i}\sigma_{y,i}}\int\limits_{x_i^L}^{\infty}dX_i\int\limits_{y_i^L}^{\infty}dY_i\phi\left(\frac{X_i-x_i}{\sigma_{x,i}}\right)\phi\left(\frac{Y_i-f(\theta,X_i)}{\sigma_{y,i}}\right),
\end{aligned}\label{eq:likelihood}
\end{equation}
where $D_X$, $U_X$ , and $L_X$ ($D_Y$, $U_Y$ and $L_Y$) mean detections and upper and lower limits in the $x$ ($y$) -axis, respectively, while $\phi$ and $\Phi$ are the probability density and cumulative density functions for the normal distribution. We solved the integrals in Eq.\,(\ref{eq:likelihood}) using the usual total least-squares approximation $f(\theta,X)=f(\theta,x_i)+f'(x_i)(X_i-x_i)$. When there are $N_{x,U}$ upper limits and $N_{x,L}$ lower limits on the $x$ -axis, we introduced the following variables:
\begin{align}
  \zeta_i^U&= \frac{x_i^U-x_i}{\sigma_{x,i}},\, i\in \{1,\cdots,N_{x,U}\}\\
  \zeta_i^L&=\frac{x_i-x_i^L}{\sigma_{x,i}},\,i\in \{1,\cdots,N_{x,L}\},
\end{align}
which are nuisance parameters that need to be estimated in the same way as the parameters $\theta$.
The final expression is
\begin{equation}
\begin{aligned}
    L&(\theta,\zeta_1^U,\cdots,\zeta_{N_{x,U}}^U, \zeta_1^L,\cdots,\zeta_{N_{x,L}}^L)=\\
    &\prod\limits_{i\in D_y}\frac{1}{\sigma_i}\phi\left(\frac{r_i}{\sigma_i}\right)\prod\limits_{i\in D_x\cap U_y}\Phi\left(\frac{r_i^U}{\sigma_i}\right)\prod\limits_{i\in D_x\cap L_y}\Phi\left(\frac{-r_i^L}{\sigma_i}\right)\\
    &\times\prod\limits_{i\in U_x\cap D_y}\Phi\left(\zeta_i^U\frac{ \sigma_i}{\sigma_{x,i}}-r_if_i'\frac{\sigma_{x,i}}{\sigma_{y,i}\sigma_i}\right)\\
    &\times\prod\limits_{i\in L_x\cap D_y}\Phi\left(\zeta_i^L\frac{ \sigma_i}{\sigma_{x,i}}+r_if_i'\frac{\sigma_{x,i}}{\sigma_{y,i}}\right)\\
   &\times\prod\limits_{i\in U_x\cap U_y}\int\limits_{-\infty}^{\frac{r_i^U}{\sigma_i}} dR\phi\left(R\right)\Phi\left(\zeta_i^U\frac{ \sigma_i}{\sigma_{x,i}}-Rf_i'\frac{\sigma_{x,i}}{\sigma_{y,i}}\right)\\
     &\times\prod\limits_{i\in U_x\cap L_y}\int\limits_{-\infty}^{-\frac{r_i^L}{\sigma_i}} dR\phi\left(R\right)\Phi\left(\zeta_i^U\frac{ \sigma_i}{\sigma_{x,i}}+Rf_i'\frac{\sigma_{x,i}}{\sigma_{y,i}}\right)\\
     &\times\prod\limits_{i\in L_x\cap U_y}\int\limits_{-\infty}^{\frac{r_i^U}{\sigma_i}} dR\phi\left(R\right)\Phi\left(\zeta_i^L\frac{ \sigma_i}{\sigma_{x,i}}+Rf_i'\frac{\sigma_{x,i}}{\sigma_{y,i}}\right)\\
     &\times\prod\limits_{i\in L_x\cap L_y}\int\limits_{-\infty}^{-\frac{r_i^L}{\sigma_i}} dR\phi\left(R\right)\Phi\left(\zeta_i^L\frac{ \sigma_i}{\sigma_{x,i}}-Rf_i'\frac{\sigma_{x,i}}{\sigma_{y,i}}\right)
\end{aligned}
\label{eq:simplified_likelihood}
,\end{equation}
where we used the following abbreviations:

\begin{figure*}
    \centering
    \includegraphics[width=\textwidth]{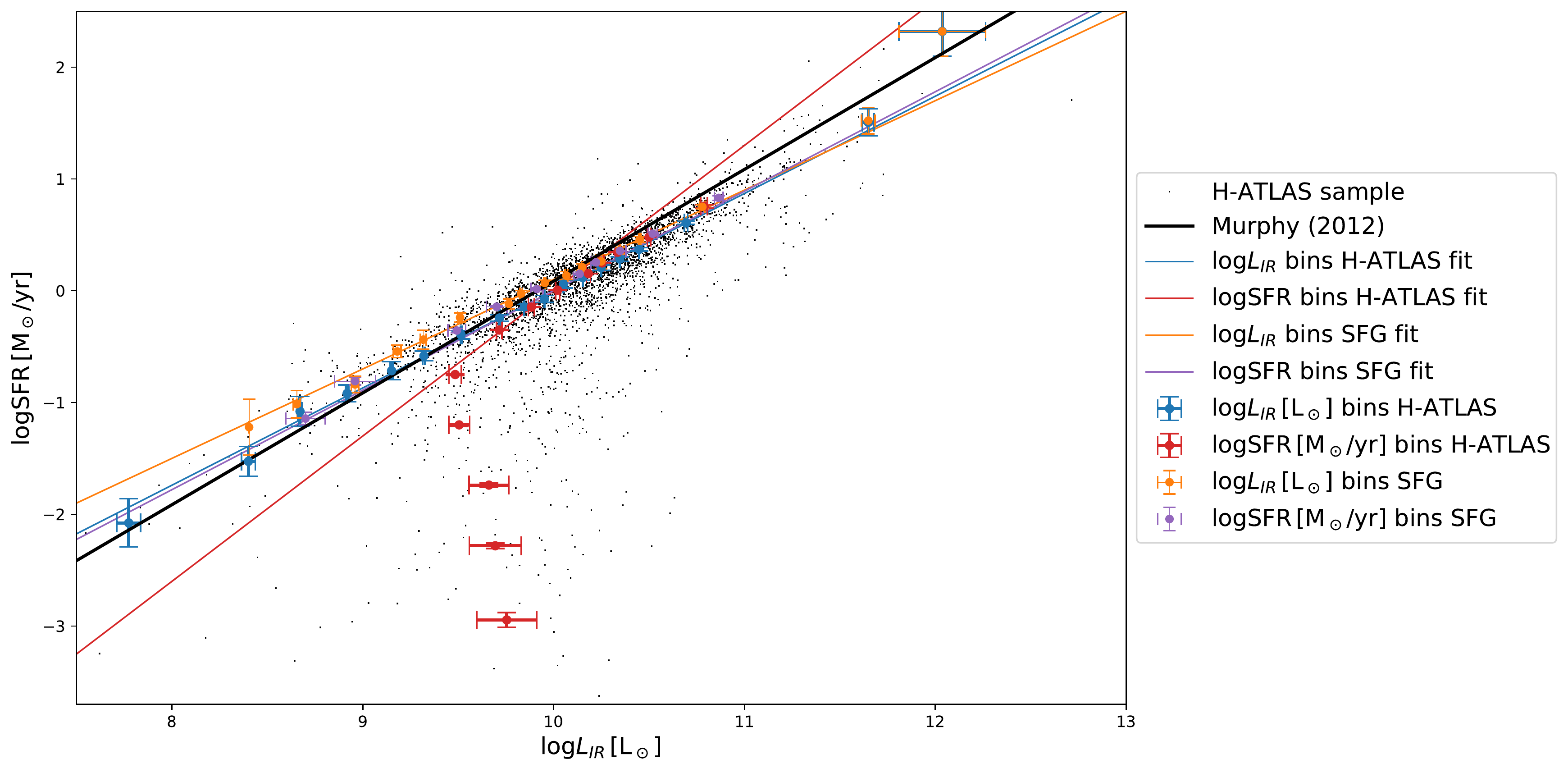}
    \caption{Relation between SFR and $L_{IR}$ for various (sub)samples,{with their respective error bars.} {The various samples are identified in the legend by the binning strategy ($L_{IR}$ and SFR bins) and by the respective sample type (H-ATLAS and SFG). {Solid lines represent the best-fitting relations for each sample, listed in Table \ref{tab:Fits}, and correspond in color to the data points. {The black dots represent  SFR and $L_{IR}$ for individual sources in the H-ATLAS sample. }}}}\label{fig:overviewb}
        \end{figure*}
\begin{figure*}
    \centering
    \includegraphics[width=\textwidth]{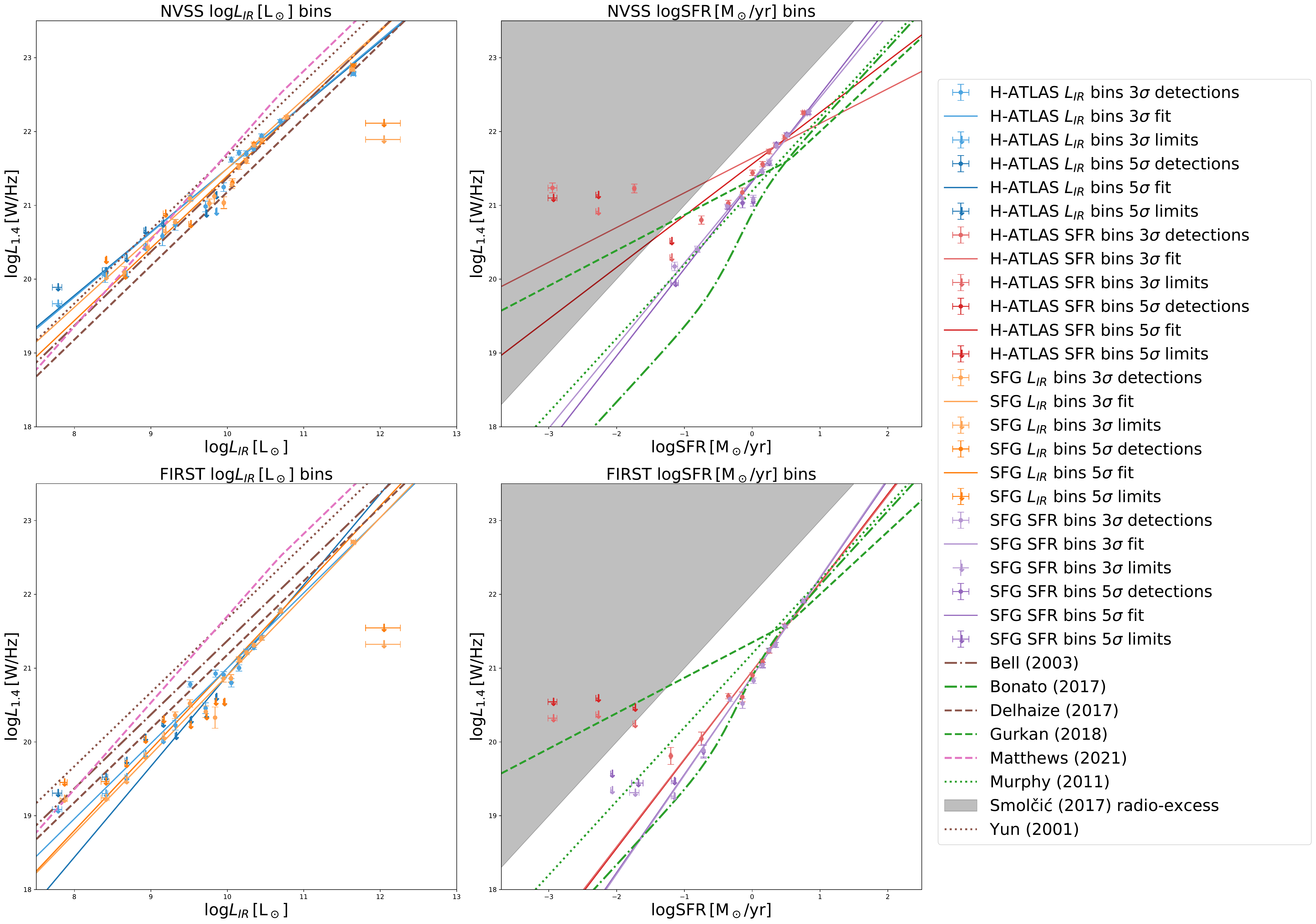}
    \caption{Radio luminosity density $\log L_{1.4}$ derived for bins of {IR} luminosity (left panels) and SFR (right panels), {with their error bars and best-fitting lines listed in Table~\ref{tab:Fits}}. {Upper limits are shown as arrows corresponding in color to detections (points).} The top row {refers to} NVSS cutouts, and the bottom row {shows} FIRST cutouts. 
Non-SFG samples, i.e., samples not cleaned using the \citet{Smolcic:17b} criterion (excluded region shown as a shaded interval), show a strong radio luminosity excess at a given SFR due to the presence of RL AGNs.}\label{fig:overview}
    \end{figure*}

\begin{table*}
\caption{Derived fits for the stacked sources in the NVSS and FIRST maps of the H-ATLAS and SFG samples.  }\label{tab:Fits}
\centering
\begin{tabular}{llllllllll}
\toprule
\toprule
Sample & Survey &  Bins & $a$ ($5\sigma$) & $a$ ($3\sigma$) &       $b$ ($5\sigma$) & $b$ ($3\sigma$) & $N_{lim}$ ($5\sigma$) &  $N_{lim}$ ($3\sigma$) \\
\midrule
\multicolumn{9}{l}{  $\log \hbox{SFR}\,[\mathrm{M_\odot/yr}]= a \log L_{\rm IR}\,[\mathrm{L_\odot}]+b$ }\\
\midrule
  H-ATLAS & Ind. & $L_{IR}$ & $0.87 \pm 0.01$ & $0.87 \pm 0.01$ & $-8.7 \pm 0.1$ & $-8.7 \pm 0.1$ &  N/A &  N/A \\
H-ATLAS &      Ind. & SFR &   $1.3 \pm 0.1$ &   $1.3 \pm 0.1$ &    $-13 \pm 1$ &    $-13 \pm 1$ &  N/A &  N/A \\
    SFG & FIRST & $L_{IR}$ & $0.80 \pm 0.02$ & $0.80 \pm 0.02$ & $-7.9 \pm 0.2$ & $-7.9 \pm 0.2$ &  N/A &  N/A \\
    SFG &      FIRST & SFR & $0.89 \pm 0.02$ & $0.90 \pm 0.02$ & $-8.9 \pm 0.2$ & $-8.9 \pm 0.2$ &  N/A &  N/A \\
    SFG & NVSS  & $L_{IR}$ & $0.82 \pm 0.02$ & $0.82 \pm 0.02$ & $-8.1 \pm 0.2$ & $-8.1 \pm 0.2$ &  N/A &  N/A \\
    SFG &      NVSS  & SFR & $0.86 \pm 0.02$ & $0.88 \pm 0.02$ & $-8.6 \pm 0.2$ & $-8.7 \pm 0.2$ &  N/A &  N/A \\
\midrule
\multicolumn{9}{l}{  $\log L_{1.4}\,\mathrm{[W/Hz]}= a \log L_{\rm IR}\,[\mathrm{L_\odot}]+b$ }\\
\midrule
   H-ATLAS & FIRST & $L_{IR}$ & $1.23 \pm 0.03$ & $1.02 \pm 0.03$ &  $8.6 \pm 0.3$ & $10.8 \pm 0.3$ & 11 (61.1 \%) & 6 (33.3 \%) \\
H-ATLAS &      FIRST & SFR & $1.35 \pm 0.04$ & $1.33 \pm 0.04$ &  $7.4 \pm 0.4$ &  $7.5 \pm 0.4$ &  3 (23.1 \%) & 3 (23.1 \%) \\
H-ATLAS & NVSS  & $L_{IR}$ & $0.86 \pm 0.02$ & $0.87 \pm 0.02$ & $12.9 \pm 0.2$ & $12.8 \pm 0.2$ &  8 (44.4 \%) & 5 (27.8 \%) \\
H-ATLAS &      NVSS  & SFR & $1.14 \pm 0.03$ & $1.15 \pm 0.04$ & $10.0 \pm 0.4$ &  $9.9 \pm 0.4$ &  3 (23.1 \%) & 2 (15.4 \%) \\
    SFG & FIRST & $L_{IR}$ & $1.10 \pm 0.03$ & $1.07 \pm 0.02$ & $10.0 \pm 0.3$ & $10.2 \pm 0.2$ & 10 (55.6 \%) & 6 (33.3 \%) \\
    SFG &      FIRST & SFR & $1.18 \pm 0.03$ & $1.21 \pm 0.04$ &  $9.1 \pm 0.3$ &  $8.8 \pm 0.4$ &  3 (25.0 \%) & 3 (25.0 \%) \\
    SFG & NVSS  & $L_{IR}$ & $0.98 \pm 0.02$ & $0.94 \pm 0.02$ & $11.6 \pm 0.2$ & $12.1 \pm 0.2$ &  5 (29.4 \%) & 3 (17.6 \%) \\
    SFG &      NVSS  & SFR & $1.05 \pm 0.04$ & $1.04 \pm 0.06$ & $10.8 \pm 0.4$ & $11.0 \pm 0.6$ &  1 (10.0 \%) &  0 (0.0 \%) \\
\midrule
\multicolumn{9}{l}{  $\log L_{1.4}\,\mathrm{[W/Hz]}= a \log \hbox{SFR}\,[\mathrm{M_\odot/yr}]+b$ }\\
\midrule
H-ATLAS & FIRST & $L_{IR}$ & $1.61 \pm 0.08$ & $1.23 \pm 0.04$ & $20.79 \pm 0.03$ & $20.97 \pm 0.02$ & 11 (61.1 \%) & 6 (33.3 \%) \\
H-ATLAS &      FIRST & SFR & $1.20 \pm 0.03$ & $1.19 \pm 0.04$ & $20.96 \pm 0.01$ & $20.96 \pm 0.02$ &  3 (23.1 \%) & 3 (23.1 \%) \\
H-ATLAS & NVSS  & $L_{IR}$ & $1.12 \pm 0.04$ & $1.11 \pm 0.04$ & $21.47 \pm 0.02$ & $21.47 \pm 0.02$ &  8 (44.4 \%) & 5 (27.8 \%) \\
H-ATLAS &      NVSS  & SFR & $0.70 \pm 0.02$ & $0.47 \pm 0.02$ & $21.56 \pm 0.01$ & $21.64 \pm 0.01$ &  3 (23.1 \%) & 2 (15.4 \%) \\
    SFG & FIRST & $L_{IR}$ & $1.49 \pm 0.05$ & $1.40 \pm 0.04$ & $20.81 \pm 0.02$ & $20.85 \pm 0.02$ & 10 (55.6 \%) & 6 (33.3 \%) \\
    SFG &      FIRST & SFR & $1.33 \pm 0.03$ & $1.34 \pm 0.03$ & $20.89 \pm 0.01$ & $20.89 \pm 0.01$ &  3 (25.0 \%) & 3 (25.0 \%) \\
    SFG & NVSS  & $L_{IR}$ & $1.27 \pm 0.05$ & $1.20 \pm 0.04$ & $21.24 \pm 0.02$ & $21.28 \pm 0.02$ &  5 (29.4 \%) & 3 (17.6 \%) \\
    SFG &      NVSS  & SFR & $1.18 \pm 0.03$ & $1.12 \pm 0.06$ & $21.32 \pm 0.02$ & $21.34 \pm 0.03$ &  1 (10.0 \%) &  0 (0.0 \%) \\
\midrule
\multicolumn{9}{l}{  $\log L_{\rm IR}\,[\mathrm{L_\odot}]-\log \hbox{SFR}\,[\mathrm{M_\odot/yr}]= a \log L_{\rm IR}\,[\mathrm{L_\odot}]+b$ }\\
\midrule
 H-ATLAS & Ind. & $L_{IR}$ & $0.13 \pm 0.01$ & $0.13 \pm 0.01$ & $8.7 \pm 0.1$ & $8.7 \pm 0.1$ &  N/A &  N/A \\
H-ATLAS &     Ind. & SFR &  $-0.2 \pm 0.1$ &  $-0.2 \pm 0.1$ &    $12 \pm 1$ &    $12 \pm 1$ &  N/A &  N/A \\
    SFG & FIRST & $L_{IR}$ & $0.21 \pm 0.02$ & $0.20 \pm 0.02$ & $7.9 \pm 0.2$ & $7.9 \pm 0.2$ &  N/A &  N/A \\
    SFG &      FIRST & SFR & $0.11 \pm 0.02$ & $0.11 \pm 0.02$ & $8.9 \pm 0.2$ & $8.9 \pm 0.2$ &  N/A &  N/A \\
    SFG & NVSS  & $L_{IR}$ & $0.18 \pm 0.02$ & $0.18 \pm 0.02$ & $8.1 \pm 0.2$ & $8.1 \pm 0.2$ &  N/A &  N/A \\
    SFG &      NVSS  & SFR & $0.14 \pm 0.02$ & $0.13 \pm 0.02$ & $8.5 \pm 0.2$ & $8.7 \pm 0.2$ &  N/A &  N/A \\
\bottomrule
\end{tabular}
\tablefoot{{For each quantity, we show the results of a two-parameter model, with parameters $a$ and $b$. When there were no limits, the orthogonal distance regression model fit was used. If limits were present in a particular dataset, we used the survival-analysis fitting strategy described in Sect. \ref{sect:survival}. The results for the H-ATLAS and SFG samples are shown in the ($5\sigma$) columns, while their variants using a BLOBCAT detection limit of $3\,\sigma$  are shown in the ($3\sigma$) columns. The $N_{\rm lim}$ column shows the total number of (x- or y-axis) nondetections for each fit. If a particular fit was not dependent on radio data, it was marked `Ind.'.}}
\end{table*}
\begin{table*}
\centering\caption{Cont.}
\ContinuedFloat
\begin{tabular}{llllllllll}
\toprule
\toprule
Sample & Survey &  Bins & $a$ ($5\sigma$) & $a$ ($3\sigma$) &       $b$ ($5\sigma$) & $b$ ($3\sigma$) & $N_{lim}$ ($5\sigma$) &  $N_{lim}$ ($3\sigma$) \\
\midrule
\multicolumn{9}{l}{  $\log L_{1.4}\,\mathrm{[W/Hz]}-\log \hbox{SFR}\,[\mathrm{M_\odot/yr}]= a \log L_{1.4}\,\mathrm{[W/Hz]}+b$ }\\
\midrule
  H-ATLAS & FIRST & $L_{IR}$ &  $0.1 \pm 0.5$ &   $0.3 \pm 0.1$ & $18 \pm 4$ & $16 \pm 5$ & 11 (61.1 \%) & 6 (33.3 \%) \\
H-ATLAS &      FIRST & SFR &  $0.1 \pm 0.4$ &   $0.1 \pm 0.3$ & $20 \pm 2$ & $19 \pm 3$ &  3 (23.1 \%) & 3 (23.1 \%) \\
H-ATLAS & NVSS  & $L_{IR}$ &  $0.2 \pm 0.5$ &   $0.2 \pm 0.3$ & $15 \pm 4$ & $18 \pm 3$ &  8 (44.4 \%) & 5 (27.8 \%) \\
H-ATLAS &      NVSS  & SFR &  $0.1 \pm 0.4$ &  $-0.2 \pm 0.4$ & $23 \pm 4$ & $27 \pm 3$ &  3 (23.1 \%) & 2 (15.4 \%) \\
    SFG & FIRST & $L_{IR}$ &  $0.3 \pm 0.2$ &   $0.3 \pm 0.2$ & $15 \pm 4$ & $15 \pm 4$ & 10 (55.6 \%) & 6 (33.3 \%) \\
    SFG &      FIRST & SFR &  $0.2 \pm 0.4$ &   $0.2 \pm 0.2$ & $17 \pm 2$ & $17 \pm 4$ &  3 (25.0 \%) & 3 (25.0 \%) \\
    SFG & NVSS  & $L_{IR}$ &  $0.3 \pm 0.4$ &   $0.2 \pm 0.3$ & $21 \pm 4$ & $19 \pm 4$ &  5 (29.4 \%) & 3 (17.6 \%) \\
    SFG &      NVSS  & SFR &  $0.2 \pm 0.3$ & $0.13 \pm 0.04$ & $17 \pm 4$ & $19 \pm 1$ &  1 (10.0 \%) &  0 (0.0 \%) \\
\midrule
\multicolumn{9}{l}{  $q_{\rm IR}= a \log L_{1.4}\,\mathrm{[W/Hz]}+b$ }\\
\midrule
   H-ATLAS & FIRST & $L_{IR}$ & $-0.3 \pm 0.5$ &   $-0.1 \pm 0.2$ & $11 \pm 5$ &  $4 \pm 5$ & 11 (61.1 \%) & 6 (33.3 \%) \\
H-ATLAS &      FIRST & SFR & $-0.3 \pm 0.3$ &   $-0.3 \pm 0.4$ &  $8 \pm 3$ &  $9 \pm 3$ &  3 (23.1 \%) & 3 (23.1 \%) \\
H-ATLAS & NVSS  & $L_{IR}$ &  $0.2 \pm 0.4$ &    $0.1 \pm 0.2$ & $-1 \pm 4$ & $-1 \pm 4$ &  8 (44.4 \%) & 5 (27.8 \%) \\
H-ATLAS &      NVSS  & SFR & $-0.2 \pm 0.4$ &   $-0.3 \pm 0.3$ &  $8 \pm 2$ &  $8 \pm 2$ &  3 (23.1 \%) & 2 (15.4 \%) \\
    SFG & FIRST & $L_{IR}$ & $-0.1 \pm 0.4$ &   $-0.1 \pm 0.3$ &  $6 \pm 2$ &  $5 \pm 2$ & 10 (55.6 \%) & 6 (33.3 \%) \\
    SFG &      FIRST & SFR & $-0.2 \pm 0.4$ &   $-0.3 \pm 0.4$ &  $8 \pm 3$ &  $8 \pm 5$ &  3 (25.0 \%) & 3 (25.0 \%) \\
    SFG & NVSS  & $L_{IR}$ &  $0.1 \pm 0.5$ &   $-0.0 \pm 0.3$ &  $1 \pm 4$ &  $1 \pm 6$ &  5 (29.4 \%) & 3 (17.6 \%) \\
    SFG &      NVSS  & SFR & $-0.1 \pm 0.4$ & $-0.05 \pm 0.05$ &  $3 \pm 4$ &  $4 \pm 1$ &  1 (10.0 \%) &  0 (0.0 \%) \\
\midrule
\multicolumn{9}{l}{  $q_{\rm IR}= a \log L_{\rm IR}\,[\mathrm{L_\odot}]+b$ }\\
\midrule
 H-ATLAS & FIRST & $L_{IR}$ & $-0.30 \pm 0.04$ &  $0.02 \pm 0.03$ &  $6.2 \pm 0.4$ & $2.8 \pm 0.3$ & 11 (61.1 \%) & 6 (33.3 \%) \\
H-ATLAS &      FIRST & SFR & $-0.33 \pm 0.04$ & $-0.34 \pm 0.03$ &  $6.4 \pm 0.4$ & $6.5 \pm 0.3$ &  3 (23.1 \%) & 3 (23.1 \%) \\
H-ATLAS & NVSS  & $L_{IR}$ &  $0.22 \pm 0.02$ &  $0.17 \pm 0.02$ &  $0.3 \pm 0.2$ & $0.8 \pm 0.2$ &  8 (44.4 \%) & 5 (27.8 \%) \\
H-ATLAS &      NVSS  & SFR & $-0.17 \pm 0.03$ & $-0.13 \pm 0.03$ &  $4.3 \pm 0.3$ & $3.9 \pm 0.3$ &  3 (23.1 \%) & 2 (15.4 \%) \\
    SFG & FIRST & $L_{IR}$ &  $0.04 \pm 0.02$ & $-0.03 \pm 0.02$ &  $2.6 \pm 0.3$ & $3.3 \pm 0.2$ & 10 (55.6 \%) & 6 (33.3 \%) \\
    SFG &      FIRST & SFR & $-0.18 \pm 0.03$ & $-0.20 \pm 0.03$ &  $4.9 \pm 0.3$ & $5.0 \pm 0.3$ &  3 (25.0 \%) & 3 (25.0 \%) \\
    SFG & NVSS  & $L_{IR}$ &  $0.14 \pm 0.02$ &  $0.10 \pm 0.02$ &  $1.2 \pm 0.2$ & $1.6 \pm 0.2$ &  5 (29.4 \%) & 3 (17.6 \%) \\
    SFG &      NVSS  & SFR & $-0.06 \pm 0.04$ & $-0.03 \pm 0.06$ &  $3.2 \pm 0.4$ & $2.9 \pm 0.6$ &  1 (10.0 \%) &  0 (0.0 \%) \\
\midrule
\multicolumn{9}{l}{  $\log L_{1.4}\,\mathrm{[W/Hz]}-\log \hbox{SFR}\,[\mathrm{M_\odot/yr}]= a q_{\rm IR}+b$ }\\
\midrule
   H-ATLAS & FIRST & $L_{IR}$ &     $-1 \pm 1$ &  $0.1 \pm 0.9$ & $25 \pm 2$ &     $23 \pm 2$ & 11 (61.1 \%) & 6 (33.3 \%) \\
H-ATLAS &      FIRST & SFR & $-0.5 \pm 0.8$ & $-0.5 \pm 0.6$ & $23 \pm 4$ &     $23 \pm 2$ &  3 (23.1 \%) & 3 (23.1 \%) \\
H-ATLAS & NVSS  & $L_{IR}$ & $-0.9 \pm 0.9$ & $-1.0 \pm 0.4$ & $24 \pm 2$ &     $24 \pm 2$ &  8 (44.4 \%) & 5 (27.8 \%) \\
H-ATLAS &      NVSS  & SFR &     $-1 \pm 1$ &     $-1 \pm 2$ & $25 \pm 3$ &     $26 \pm 4$ &  3 (23.1 \%) & 2 (15.4 \%) \\
    SFG & FIRST & $L_{IR}$ &     $-1 \pm 1$ &     $-1 \pm 1$ & $25 \pm 4$ &     $24 \pm 3$ & 10 (55.6 \%) & 6 (33.3 \%) \\
    SFG &      FIRST & SFR & $-0.9 \pm 0.8$ &     $-1 \pm 1$ & $24 \pm 2$ &     $24 \pm 2$ &  3 (25.0 \%) & 3 (25.0 \%) \\
    SFG & NVSS  & $L_{IR}$ & $-0.2 \pm 0.9$ &     $-1 \pm 1$ & $23 \pm 3$ &     $24 \pm 2$ &  5 (29.4 \%) & 3 (17.6 \%) \\
    SFG &      NVSS  & SFR &     $-0 \pm 1$ & $-1.6 \pm 0.4$ & $24 \pm 5$ & $25.4 \pm 0.9$ &  1 (10.0 \%) &  0 (0.0 \%) \\
\bottomrule
\end{tabular}
\end{table*}

\begin{align}
\hat x_i&=\begin{cases}x_i & \mathrm{detections}\\
    x_i^U-\sigma_{x,i}\zeta_i^U & \mathrm{upper\,limits}\\
    x_i^L+\sigma_{x,i}\zeta_i^L&\mathrm{lower\,limits}
    \end{cases}\\
f_i & = f(\theta,\hat x_i)\\
f'_i & = f'(\theta,\hat x_i)\\
r_i&=y_i-f_i\\
r_i^U&=y_i^U-f_i\\
r_i^L&=y_i^L-f_i\\
\sigma_i&=\sqrt{\sigma_{y,i}^2+(f'_i\sigma_{x,i})^2}.
\end{align}
We maximized the likelihood in Eq.\,(\ref{eq:simplified_likelihood}) using the Markov chain Monte Carlo {(MCMC)} method implemented in the package emcee \citep{Foreman13}. We started by giving uniform priors for $\theta$ and $\zeta$s. Because $\zeta$s are just normalized $x$ values that are expected to be mostly positive by definition, we used a fixed range $\zeta_i\in\langle -1,10]$.

{We tested the survival analysis codes using Monte Carlo simulations to fit the linear model. We made {5000} simulations for samples of sizes in the range from 5 to 15 data points, chosen to encompass the number of bins in the real dataset. We chose to simulate linear models with both slopes and intercepts in the intervals $[-2,2]$. The simulated $y$-values were centered on the chosen linear model, with deviations drawn from the normal distribution $\mathcal{N}(0,\sigma^2)$, where $\sigma$ was chosen to vary in the range $[10^{-4},10^{-1}]$. The error on $\sigma_x$ was set equal to $\sigma$. Furthermore, the probability of detection, $p_D$, was chosen to be between 0.4 and 1. The probability that a point is an upper limit in each axis was chosen from a uniform distribution in the interval $[0,1-p_D]$, and the probability for a data point to be a lower limit on an axis was computed so that the probabilities summed to one. The labels of a detection, upper and lower limit were then drawn for each data point and for each axis from these distributions. If a point was an upper limit in the x (y) axis, its value was shifted upwards by $5\sigma_x$ ($5\sigma_y$). Conversely, if it was a lower limit in the x (y) axis, its value was shifted downwards by $5\sigma_x$ ($5\sigma_y$).}

{The differences between the slopes and intercepts obtained from simulations and the input values are shown in Fig.\,\ref{fig:FigureSimulationDiff} as functions of the ratio of {the number of} lower and upper limits on both axes ($R_{x,L}$, $R_{x,U}$, $R_{y,L}$, $R_{x,L}$), of $\log\sigma$ and of the size of the sample. No systematic offsets were found in these distributions. We also quantified this finding by computing the P-value in each bin of $R_{x,L}$, $R_{x,U}$, $R_{y,L}$, $R_{x,L}$, $\log\sigma$ , and size. To compute the P-values, we first computed the standard Z-scores {(i.e., the difference between the input and simulated value divided by the standard deviation of the simulations)} of the differences between simulated, $a$, and input slopes, $\hat a$, as well as those between simulated, $b$, and input intercepts, $\hat b$. Standard deviations were computed from the MCMC-derived covariance matrix. The P-values of each estimate in each bin were then combined using the Fisher method \citep[see, e.g., ][]{fisher1925statistical}. The resulting P-values, shown in Fig. \ref{fig:FigureSimulationPvalues}, suggest that there is no significant offset in any bin below $\log\sigma\sim{-1.5}$ at a significance value of $5\%$, {which is comparable to the $\log\sigma$-s in the (sub)sample (ranging from $-1.6$ to $-1.2$). By considering the derived parameter differences and $Z$-scores, we conclude that there is an expected bias in the derived parameters, but that the bias is lower than the parameter errors and therefore is not large enough to be relevant for our analysis.}}

\section{Results}\label{sect:results}

\begin{figure*}
    \centering

    \includegraphics[width=\textwidth]{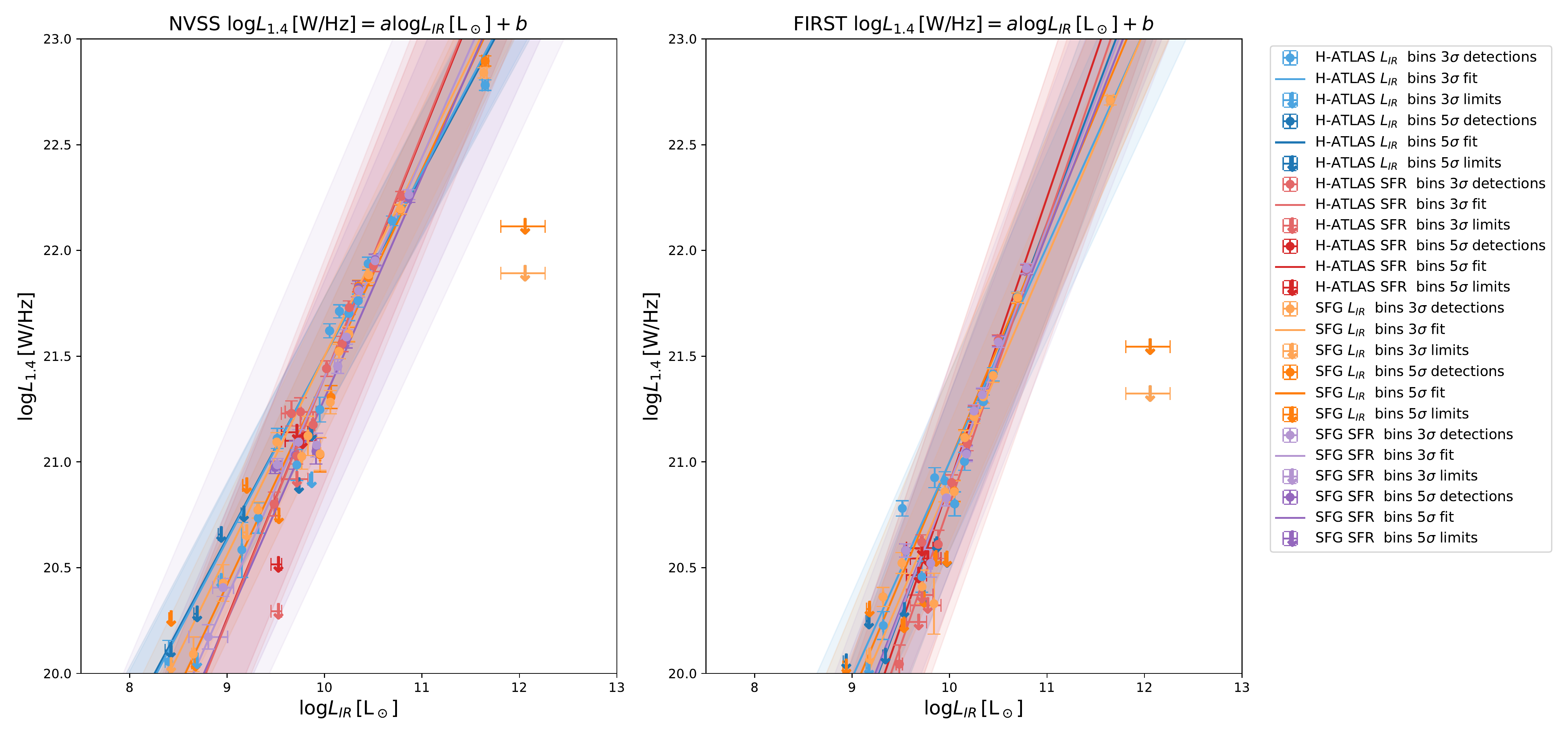}
    
    \includegraphics[width=\textwidth]{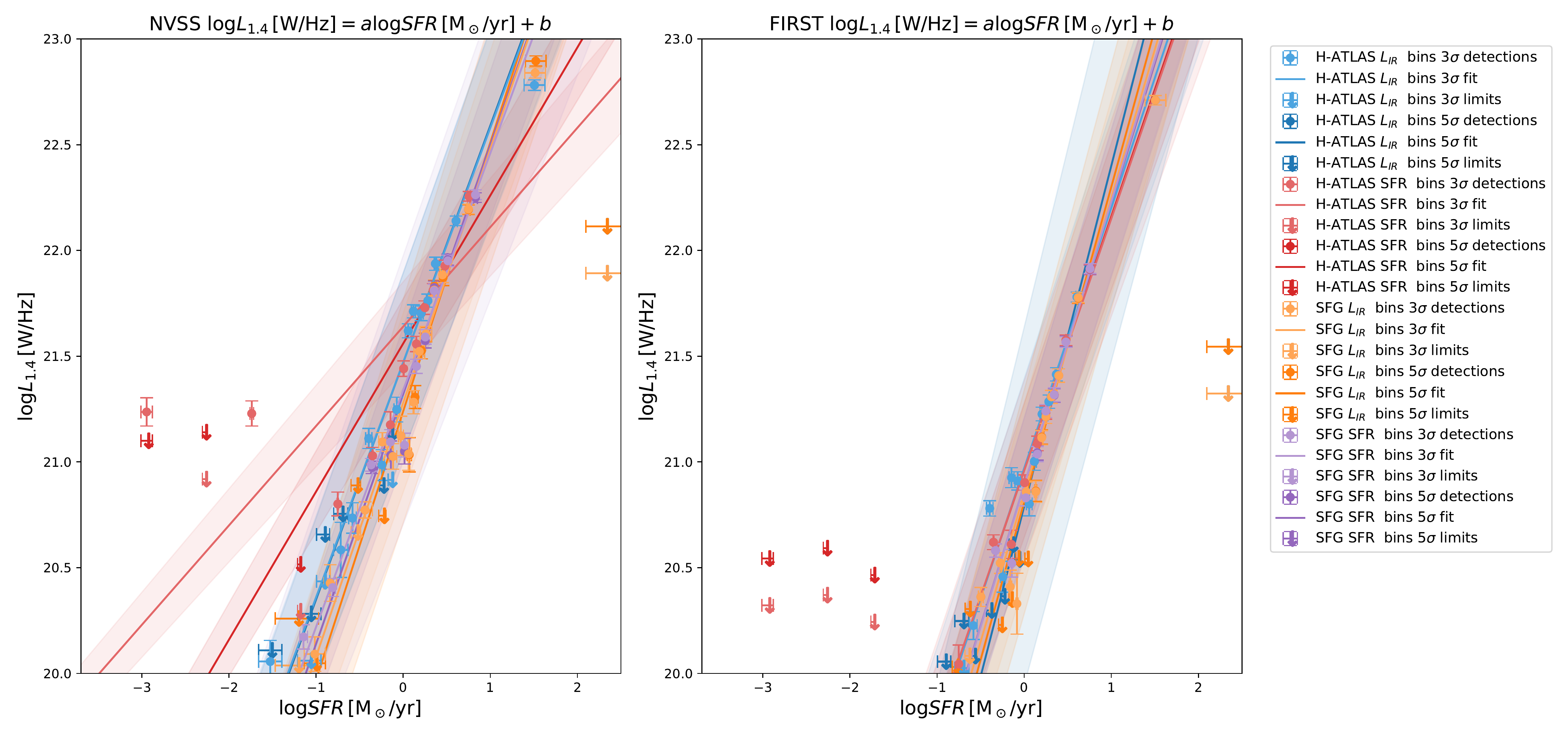}
\caption{{Best-fitting lines from Table \ref{tab:Fits} for which the relative errors on the two parameters $a$ and $b$ are lower than $20\%$. {Solid lines represent the best-fitting relations for each sample, listed in Table \ref{tab:Fits}, while their corresponding confidence intervals are shown by the shaded bands. {The left panels show results for the NVSS cutouts, and the right panels show the results for the FIRST cutouts.} Detections are marked with points, and arrows are used to show nondetections, both with accompanying error bars on both axes.}}}
   \label{fig:SampleFits}
\end{figure*}
\begin{figure*}
    \centering\ContinuedFloat
    \includegraphics[width=\textwidth]{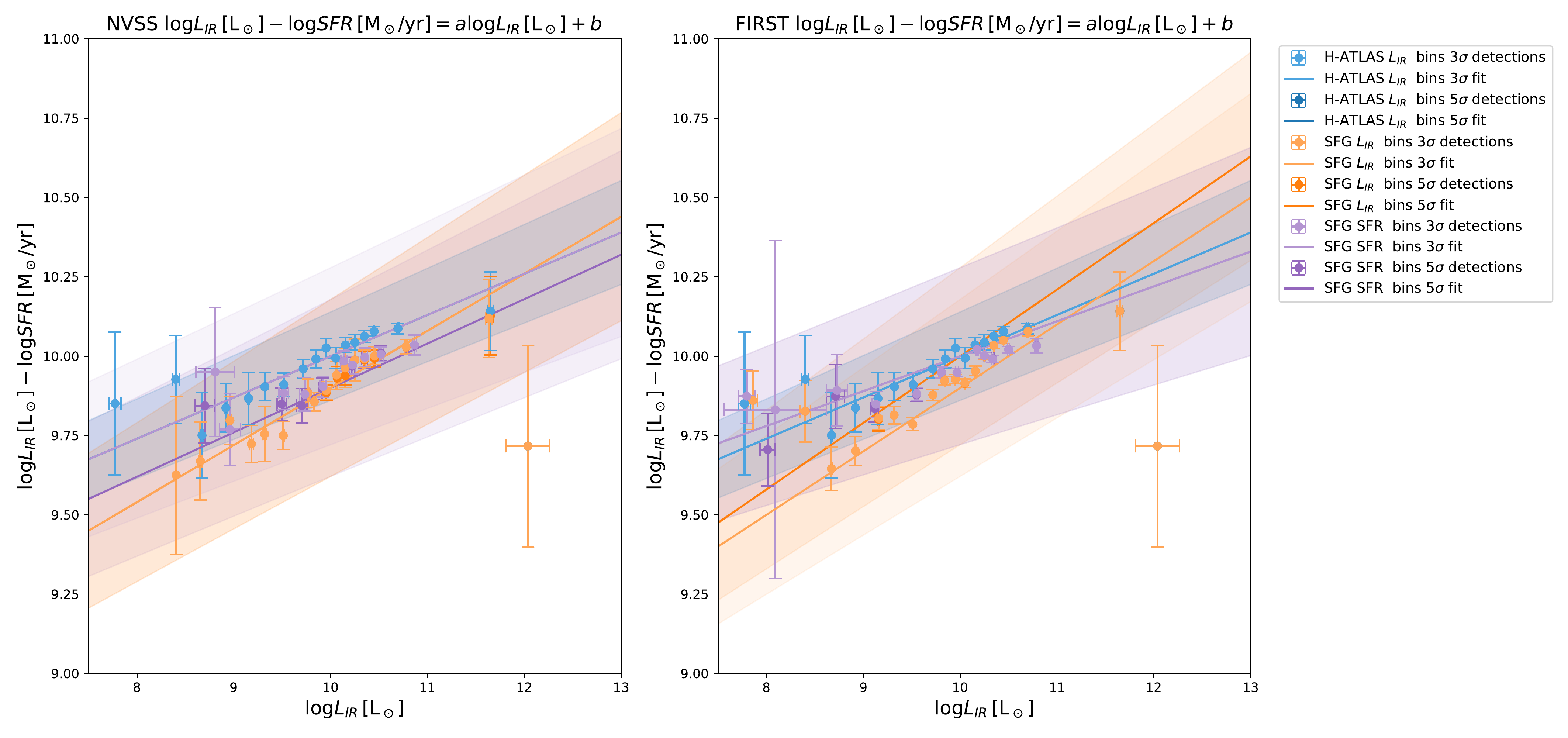}
    
    \includegraphics[width=\textwidth]{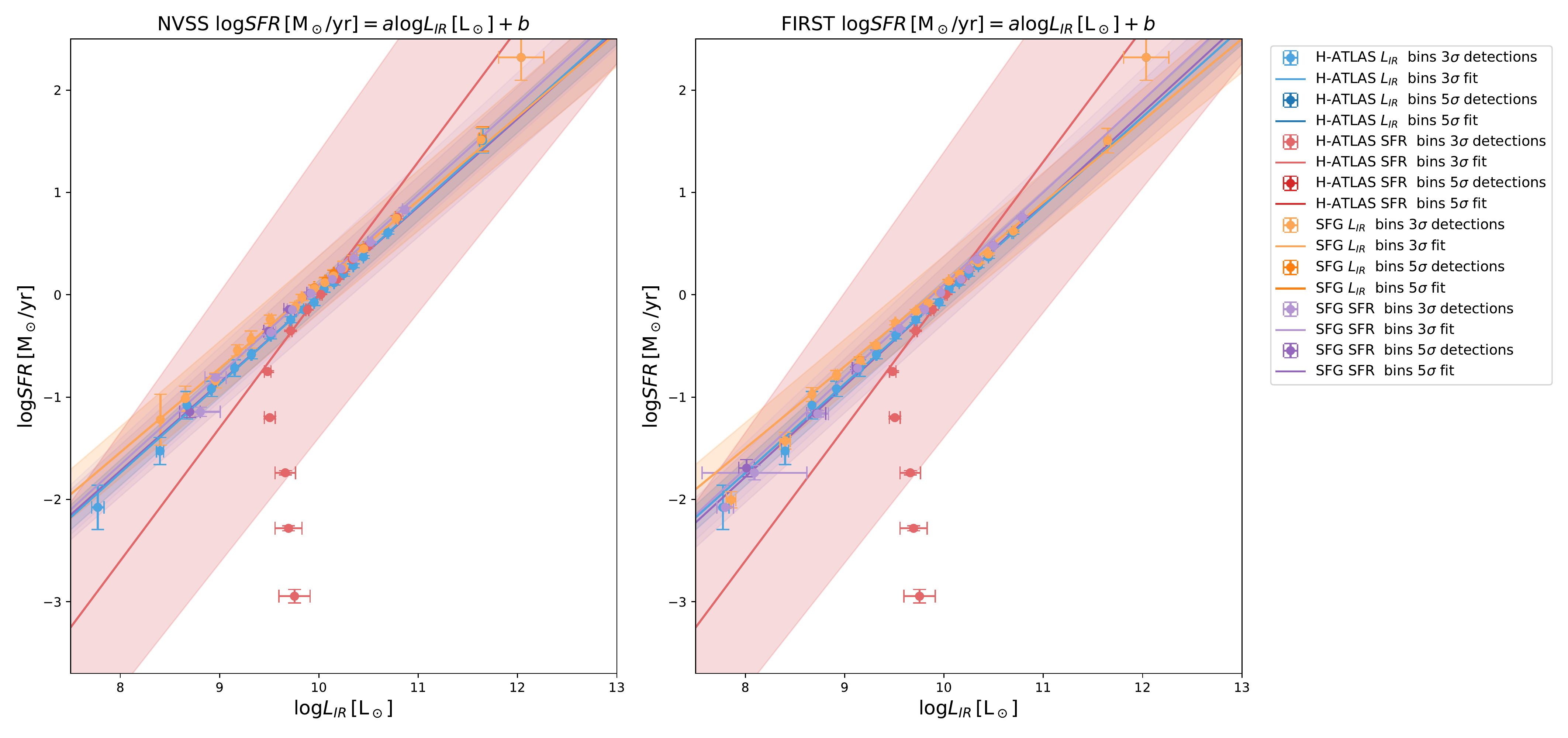}
    \caption{Cont.}

\end{figure*}
    \begin{figure*}
    \centering\ContinuedFloat
    \includegraphics[width=\textwidth]{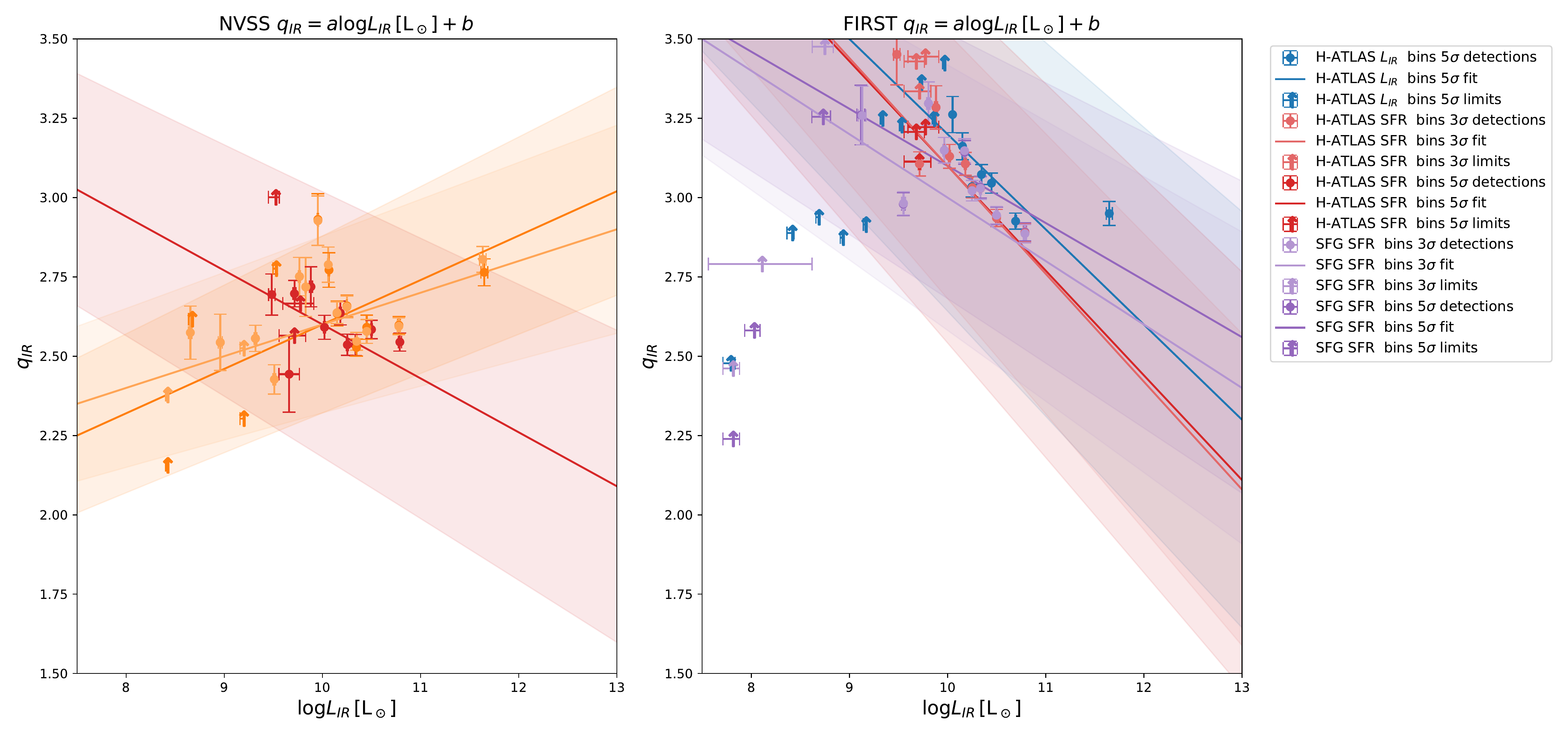}
    \caption{Cont.}

\end{figure*}

We stacked the NVSS and FIRST map cutouts using $L_{\rm IR}$ and SFR bins, as detailed in Table\,\ref{tab:Number_counts_table}. We separately considered the H-ATLAS sample, the SFG subsample, the H-ATLAS $3\sigma$ sample, and the SFG $3\sigma$ subsample. 

The {IR} luminosity-SFR relation is shown in Fig. \ref{fig:overviewb}, and the radio luminosity densities for the different {IR} luminosity and SFR bins {are shown} in Fig. \ref{fig:overview}. For each bin, the radio luminosity density was derived using BLOBCAT on the stacked cutout image. If no signal was detected, a $5\sigma$ ($3\sigma$ for the extended sample) limit was computed from the RMS noise in the stacked cutout.

The results of the stacking procedure and of survival analysis can be found in Table\,\ref{tab:Fits}. In this table we show the results of our fits. When there were no upper or lower limits on either axis, an orthogonal distance regression model fit was used for the linear model, while survival analysis was used when limits were present on either axis. The results are plotted in Fig. \ref{fig:SampleFits} for a subset of fits for which the relative errors of the two linear-model parameters were lower than $20\%$. 

 We tested linear relation fits connecting $\log\mathrm{SFR}$ and the logarithm of the {IR} luminosity, $\log{L_{IR}}$. We tested for the significance of nonlinearities in the  $\log{L_{IR}}-\log\mathrm{SFR}$ relation by fitting a linear relation to the transformed value, $\log{L_{IR}}-\log\mathrm{SFR}$ as a function of {IR} luminosity, which tests for the presence of a quadratic ($\log{L_{IR}}$) term in the $\log{L_{IR}}-\log\mathrm{SFR}$ dataset. We applied the same fitting strategy to the radio luminosity, exploring its dependence on SFR and {IR} luminosity, and tested for possible nonlinearities by fitting a linear relation to the transformed value, $\log{L_{1.4}}-\log\mathrm{SFR}$, as a function of $\log{L_{1.4}}$. We additionally verified whether the $q_{IR}$ parameter depends on {IR} and radio luminosity, or if it can affect the $\log{L_{1.4}}-\log\mathrm{SFR}$ value.
 
\section{Discussion}\label{sect:discussion}

{We chose to compare our results with a selection of local studies and studies with a wide redshift range at or converted to 1.4\,GHz. This is representative of the published relations (see Sect.~\ref{sect:introduction}). \citet{Yun01} observed an IRAS galaxy sample cross-matched to the NVSS survey in a similar redshift range as our study, {while \citet{Bell2003EstimatingCorrelation} assembled a diverse sample of galaxies with accompanying {far ultraviolet}, optical, IR, and radio luminosities to explore the infrared-radio correlation. } 
  \citet{Murphy2011CalibratingNGC6946} analyzed Ka-band observations of star-forming regions in NGC 6946, which they expanded to a sample of 56 nearby galaxies at $33$ GHz with a median beam width of $25\,\mathrm{arcsec}$ in \citet{Murphy2012TheRegions}. \citet{Delhaize17} studied the $q$ parameter out to $z\sim 5$ using a survival analysis technique to account for {IR} and $3\,\mathrm{GHz}$ radio nondetections, while \citet{Bonato2017} studied the radio luminosity function of {SFGs} and searched for nonlinearity in the relation of radio luminosity to SFR out to $z\sim 5$. \cite{Gurkan2018LOFAR/H-ATLAS:Relation} studied the relation between LOFAR and H-ATLAS observations of SFGs, while \citet{Matthews2021} studied a large, local, volume-limited sample of SFGs selected at 1.4 GHz.}

{Figure\,\ref{fig:overviewb} ($L_{\rm IR}$ versus SFR) shows consistency with the \citet{Murphy2012TheRegions} linear relation between $L_{\rm IR}$ and SFR for both SFG samples with regard to the slope, {which they derived using the $33\,\mathrm{GHz}$ data and a \citet{Kroupa2001} initial mass function.}  We also fit the $\log L_{\mathrm{IR}}/\mathrm{SFR}$ ratio versus $L_{\mathrm{IR}}$ with a linear model. {We find a small positive slope at a significance of $5\sigma$  for all samples and binning strategies, except for the SFR bins using the H-ATLAS sample, for which the error on the slope was too large for us to accurately determine the slope. Overall, using SFR bins yielded a smaller slope ($\sim 0.1$), while the $L_{IR}$ binning produced a slope of $\sim0.2$. In both cases, however, the data indicate higher $L_{\mathrm{IR}}/\mathrm{SFR}$ ratios for higher {IR} luminosity, generally corresponding to higher obscuration by dust. This suggests that the MAGPHYS fits may either slightly overestimate $L_{\rm IR}$ or underestimate the SFR for the more obscured galaxies. }}

In Fig.\,\ref{fig:overview} we show the correlations between the radio luminosity density, $L_{1.4}$, derived from the NVSS and FIRST maps, and $L_{\rm IR}$ or SFR.  {The NVSS radio luminosities, indicated by the best-fitting lines in the left panels of Fig.\,\ref{fig:overview},} are substantially higher than the FIRST radio luminosities, which is consistent with extended emission of our low-$z$ galaxies being missed at the FIRST resolution, as pointed out by \citet{Jarvis2010}. Therefore we focused on NVSS luminosities. Our results for the SFG samples are consistent with the linear relation between $L_{1.4}$ and $L_{\rm IR}$ by \citet{Bell2003EstimatingCorrelation}. The relations by \citet{Yun01} and by \citet{Delhaize17} are somewhat higher and somewhat lower than our mean relation, respectively. {Our results do not support the sublinear relation between $L_{1.4}$ and $L_{\rm IR}$ reported by \citet{Gurkan2018LOFAR/H-ATLAS:Relation}, nor the substantially superlinear relation by \citet{Matthews2021}.}

As expected, H-ATLAS sample {data points}, which also contain RL AGNs, show substantial deviations from a linear relation at low-IR luminosities. {The fitted slope of the $\log L_{1.4} - \log L_{\rm IR}$ relation based on NVSS data is sublinear at the $7\,\sigma$ level. A similar trend, with an even more significant sublinearity, is seen in the $\log L_{1.4}-\log \hbox{SFR}$ relation for NVSS flux densities and SFR bins. On the other hand, in the other cases, the $\log L_{1.4} - \log L_{\rm IR}$ relation is superlinear. In all cases, the quality of the fit is quite poor. This complex situation arises because of the contaminating effect of RL AGNs, whose radio emission dominates especially at low and high SFRs or $L_{\rm IR}$s, flattening or steepening the relations, respectively. Moreover, local radio AGNs are generally hosted by passive early-type galaxies whose {IR} luminosity may be mostly due to dust heated by old stellar populations rather than to star formation. }

We find that the upper limits do not significantly affect the trends when they are accounted for using survival analysis. The simulations suggest that the datasets with a higher percentage of upper limits should be progressively less reliable, but large systematic offsets in the derived linear model parameters are not expected. The $\log L_{\mathrm{1.4}}/\mathrm{SFR}$ ratio fits do not point toward a statistically significant nonlinearity, even when the upper limits are accounted for. We therefore conclude that the survival analysis results are consistent with the linear $L_{1.4}$--SFR relation by \citet{Murphy2011CalibratingNGC6946}. There is no evidence of a stronger decline of the $L_{1.4}$/SFR ratio at low luminosity that was obtained based on one of the relations derived by \citet{Bonato2017} from the comparison of the local SFR function with the local 1.4\,GHz luminosity function. 

{The $q_{\rm IR}$ does not show any statistically significant dependence on radio luminosity. The dependence on $L_{\rm IR}$ in the case of the H-ATLAS samples is the trivial consequence of the fact that the radio luminosity of RL AGNs is independent of the SFR, which enters the definition of $q_{\rm IR}$ and is correlated to $L_{\rm IR}$.  The $L_{1.4}/\mathrm{SFR}$ ratio shows no statistically significant correlation with $q_{\rm IR}$.} We find no statistically significant differences between the results for the H-ATLAS and H-ATLAS--$3\sigma$ samples and for the SFG and SFG--$3\sigma$ samples. 

{The radio-excess criterion is not perfect in discerning AGNs from {SFGs}. \citet{Smolcic:17b} defined the moderate-to-high radiative luminosity AGN (HLAGN) by a combination of X-ray, MIR   color--color,   and   SED-fitting criteria. The low-to-moderate radiative luminosity AGN (MLAGN) are defined by either a color selection criterion combined with the absence of \textit{Herschel} detections (quiescent MLAGN) or by the $3\sigma$ radio excess. Because we use a \textit{Herschel}--selected sample, we do not expect contamination by radio-quiet MLAGN, and the radio-excess criterion eliminates the presence of radio-excess MLAGN. The contamination by HLAGN is harder to remove because only about $30\%$ of HLAGN exhibit radio excess. In Fig. \ref{fig:overview} we show the region that was excluded in the definition of the SFG subsamples as a shaded interval. We find that the only H-ATLAS sample bins that are significantly affected by this cut are those with a lower SFR. \citet{Delhaize17} found that the HLAGN have a lower $q_{IR}$ value than {SFGs}, which could in part explain the large dispersion in the fitting parameters for our $q_{IR}$ fits.}

\section{Summary and conclusions}\label{sect:conclusions}
We have investigated the IR-radio and SFR-radio correlations for a sample that is complete to over 91\%  of {3,328} $z\le 0.1$ galaxies (H-ATLAS sample). All galaxies have spectroscopic redshifts. The sample was drawn from the H-ATLAS catalog, which covers three equatorial fields that encompass a total area of $161.6\,\hbox{deg}^2$. The wealth of multifrequency data that is available for these sources has allowed us to determine total {IR} (8--$1000\,\mu$m) luminosities and SFR for all sources using the MAGPHYS package \citep{daCunha2008}.

Radio data were provided by the FIRST and NVSS surveys. We extracted 1.4\,GHz flux densities from radio images and checked for possible biases. The FIRST flux densities are systematically lower than those of the NVSS, as expected because the FIRST 5 arcsec beam misses a substantial fraction of extended emission at the low redshifts of our sources. We therefore focused on NVSS flux densities. 

An SFG subsample was built by removing from the H-ATLAS sample all sources showing a radio excess according to the \citet{Smolcic:17b} criterion.  However, only a minor fraction of sources are above the $5\,\sigma$ detection threshold: {81 FIRST (104 NVSS)} SFG detections out of a total of {84 FIRST (138 NVSS)} detections. Sources below the threshold were binned in $\log(L_{\rm IR}),$ and for each bin, we performed a stacking analysis on the NVSS and FIRST images. {Statistical $5\,\sigma$ detections were obtained for most  bins, 7 (10) out of 18 FIRST (NVSS) $\log L_{\rm IR}$ bins and 10 (10) out of 13 FIRST (NVSS) $\log \hbox{SFR}$ bins.} For bins lacking detection, we extended the \citet{Sawicki2012SEDfit:Data} survival analysis technique.

Our sample has allowed us to investigate the $L_{\rm IR}$--SFR, the $L_{1.4}$--$L_{\rm IR}$ and the $L_{1.4}$--SFR correlations. In all cases, the H-ATLAS sample showed  much larger dispersions than the SFG sample, betraying the presence of a substantial contamination by radio AGNs. This contamination is stronger at {the highest and at the lowest radio luminosities and tends to yield superlinear or sublinear $L_{1.4}$ versus $L_{\rm IR}$ and $L_{1.4}$ versus SFR relations, respectively}. To better test low luminosities and low SFRs, we also considered $3\,\sigma$ detections. The results from the $5\,\sigma$ and $3\,\sigma$ samples are fully consistent with each other. {The SFG samples have allowed us to probe the ranges $7.5\simlt \log(L_{\rm IR}/L_\odot)\simlt 13$, $-3.5\simlt \log(\hbox{SFR}/M_\odot\,\hbox{yr}^{-1})\simlt 2.5$ and $20\simlt \log(L_{1.4}/\hbox{W}\,\hbox{Hz}^{-1})\simlt 22$.} In these low-luminosity {or} low-SFR ranges, deviations from linearity are expected to show up more clearly. 

Our results are fully consistent with the linear relations between $L_{\rm IR}$ and SFR by \citet{Murphy2012TheRegions}, between $L_{1.4}$ and $L_{\rm IR}$ by \citet{Bell2003EstimatingCorrelation}, and  between $L_{1.4}$ and SFR by \citet{Murphy2011CalibratingNGC6946}. {However, our data are also consistent with the small deviations from linearity that were reported by some studies}. {The normalization of our} $L_{1.4}$ versus $L_{\rm IR}$ relation is intermediate between those by \citet{Yun01}, which is somewhat higher, and by \citet{Delhaize17}, which is somewhat lower.

\section*{Acknowledgements}
{This work is financed within the Tenure Track Pilot Program of the
Croatian Science Foundation and the \'Ecole Polytechnique F\'ed\'erale de
Lausanne and the Project TTP-2018-07-1171 Mining the Variable Sky, with
funds of the Croatian-Swiss Research Program.}
Amirnezam Amiri thanks Habib Khosroshahi (IPM), Ali Dariush (Cambridge University), and Mohammad H. Zhoolideh Haghighi (IPM) for useful discussions. Kre\v{s}imir Tisani\'{c} thanks Mladen Novak (Max-Planck-Institut f\"{u}r Astronomie).

\bibliographystyle{aa}
\bibliography{references}


\end{document}